\documentclass{vldb}
\usepackage{graphicx}
\usepackage{balance}  
\usepackage{subcaption}
\usepackage{tabularx}
\usepackage{array}
\usepackage{booktabs}
\usepackage{rotating}
\usepackage{cite}
\usepackage{url}
\usepackage{enumitem}
\usepackage{float}
\usepackage{placeins}
\usepackage{xspace}
\usepackage{hyperref}

\newcommand{\sparagraph}[2][.]{\vspace{1mm}\noindent {\bf #2#1}}

\newcommand{\books}{\texttt{amzn}\xspace}
\newcommand{\face}{\fb}
\newcommand{\fb}{\texttt{face}\xspace}
\newcommand{\osm}{\texttt{osm}\xspace}
\newcommand{\wiki}{\texttt{wiki}\xspace}

\usepackage{xcolor}

\begin{document}


\title{Benchmarking Learned Indexes}



%
%
%
%

\numberofauthors{1} 

\author{
\alignauthor
Ryan Marcus$^{13}$, Andreas Kipf$^{1}$, Alexander van Renen$^{2}$, Mihail Stoian$^{2}$,\\Sanchit Misra$^{3}$, Alfons Kemper$^{2}$, Thomas Neumann$^{2}$, Tim Kraska$^{1}$\\
\vspace{.4em}
\affaddr{$^1$MIT CSAIL \ \ \ \ $^2$TUM \ \ \ \  $^3$Intel Labs}\\
\vspace{.2em}
{\small \texttt{\{ryanmarcus, kipf, kraska\}@mit.edu \ \{renen, stoian, kemper, neumann\}@in.tum.de \ sanchit.misra@intel.com}}
}


\maketitle

\begin{abstract}
Recent advancements in learned index structures propose replacing existing index structures, like B-Trees, with approximate learned models. In this work, we present a unified benchmark that compares well-tuned implementations of three learned index structures against several state-of-the-art ``traditional'' baselines. Using four real-world datasets, we demonstrate that learned index structures can indeed outperform non-learned indexes in read-only in-memory workloads over a dense array. We also investigate the impact of caching, pipelining, dataset size, and key size. We study the performance profile of learned index structures, and build an explanation for why learned models achieve such good performance. Finally, we investigate other important properties of learned index structures, such as their performance in multi-threaded systems and their build times.
\end{abstract}

\begin{sloppypar}
\section{Introduction}
While index structures are one of the most well-studied components of database management systems, recent work~\cite{ml_index, lis-survey} provided a new perspective on this decades-old topic, showing how  machine learning techniques can be used to develop so-called \emph{learned} index structures. Unlike their traditional counterparts (e.g.,~\cite{art,fast,ibtree,surf,wormhole,hot}), learned index structures build an explicit model of the underlying data to provide effective indexing. 

Since learned index structures have been proposed, they have been criticized~\cite{caseforbtree,stanfordblog}.
The main reasons for these criticisms were the lack of an efficient open-source implementation of the learned index structure, inadequate data-sets, and the lack of a standardized benchmark suite to ensure a fair comparison between the different approaches. 

Even worse, the lack of an open-source implementation forced researchers to  re-implement the techniques of~\cite{ml_index}, or only use back-of-the-envelop calculations, to compare against learned index structures. 
While not a bad thing \emph{per se}, it is easy to leave the baseline unoptimized, or make other unrealistic assumptions, even with the best of intentions, potentially rendering the main takeaways void.

For example, recently Ferragina and Vinciguerra proposed the PGM index~\cite{pgm-index}, a learned index structure with interesting theoretical properties, which is recursively built bottom-up. Their experimental evaluation showed that the PGM-index was strictly superior to traditional indexes as well as their own implementation of the original learned index~\cite{ml_index}. This strong result surprised the authors of \cite{ml_index}, who had experimented with bottom-up approaches and usually found them to be slower to execute (see Section~\ref{sec:lis_discussion} for a discussion why this may be case). This motivated us to investigate if the results of~\cite{pgm-index} would hold against tuned implementations of the original learned index~\cite{ml_index} and other structures.

Further complicating matters, learned structures have an ``unfair'' advantage on synthetic datasets, as synthetic datasets are often surprisingly easy to learn. 
Hence, it is often easy to show that a learned structure outperforms the more traditional approaches just by using the right kind of data. 
While this is also true for almost any benchmark, it is much more pronounced for learned algorithms and data structures as their entire goal is to automatically adjust to the data distribution and even the workload. 

In this paper, we try to address these problems on three fronts: (1) we provide a first open-source implementation of RMIs for researchers to compare against and improve upon, (2) we created a repository of several real-world datasets and workloads for testing, and (3) we created a benchmarking suite, which makes it easy to compare against learned and traditional index structures. 
To avoid comparing against weak baselines, our open-source benchmarking suite~\cite{url-sosd} contains implementations of index structures that are either widespread, tuned by their original authors, or both. 

\sparagraph{Understanding learned indexes} In addition to providing an open source benchmark for use in future research, we also tried to achieve a deeper understanding of learned index structures, extending the work of~\cite{sosd}. 

First, we present a Pareto analysis of three recent learned index structures (RMIs~\cite{ml_index}, PGM indexes~\cite{pgm-index}, and RS indexes~\cite{radix-spline}) and several traditional index structures, including trees, tries, and hash tables. We show that, in a warm-cache, tight-loop setting, all three variants of learned index structures can provide better performance/size tradeoffs than several state-of-the-art traditional index structures. We extend this analysis to multiple dataset sizes, 32 and 64-bit integers, and different search techniques (i.e., binary search, linear search, interpolation search).

Second, we analyze \emph{why} learned index structures achieve such good performance. While we were unable to find a single metric that fully explains the performance of an index structure (it seems intuitive that such a metric does not exist), we offer a statistical analysis of performance counters and other properties. The single most important explanatory variable was cache misses, although cache misses alone are not enough for a statistically significant explanation. Surprisingly, we found that branch misses do \emph{not} explain why learned index structures perform better than traditional structures, as originally claimed in~\cite{ml_index}. In fact, we found that both learned index structures and traditional index structures use branching efficiently.

Third, we analyze the performance of a wide range of index structures in the presence of memory fences, cold caches, and multi-threaded environments, to test their behavior under more realistic settings. In all scenarios, we found that learned approaches perform surprisingly well. 

However, our study is not without its limitations.  We focused only on read-only workloads, and we tested each index structure in isolation (e.g., a lookup loop, not with integration into any broader application).
While this certainly does not cover all potential use cases, in-memory performance is increasingly important, and many write-heavy DBMS architectures are also moving towards immutable read-only data-structures (for example, see LSM-trees in RocksDB~\cite{lsm_survey,url-rocksdb}). Hence, we believe our benchmark can still guide the design of many systems to come and, more importantly, serve as a foundation to develop benchmarks for mixed read/write workloads and the next generation of learned index structures which supports writes~\cite{alex,pgm-index,fiting_tree}. 




\section{Formulation \& definitions}
\label{sec:formulation}

\begin{figure}
  \centering
  \includegraphics[width=0.9\linewidth]{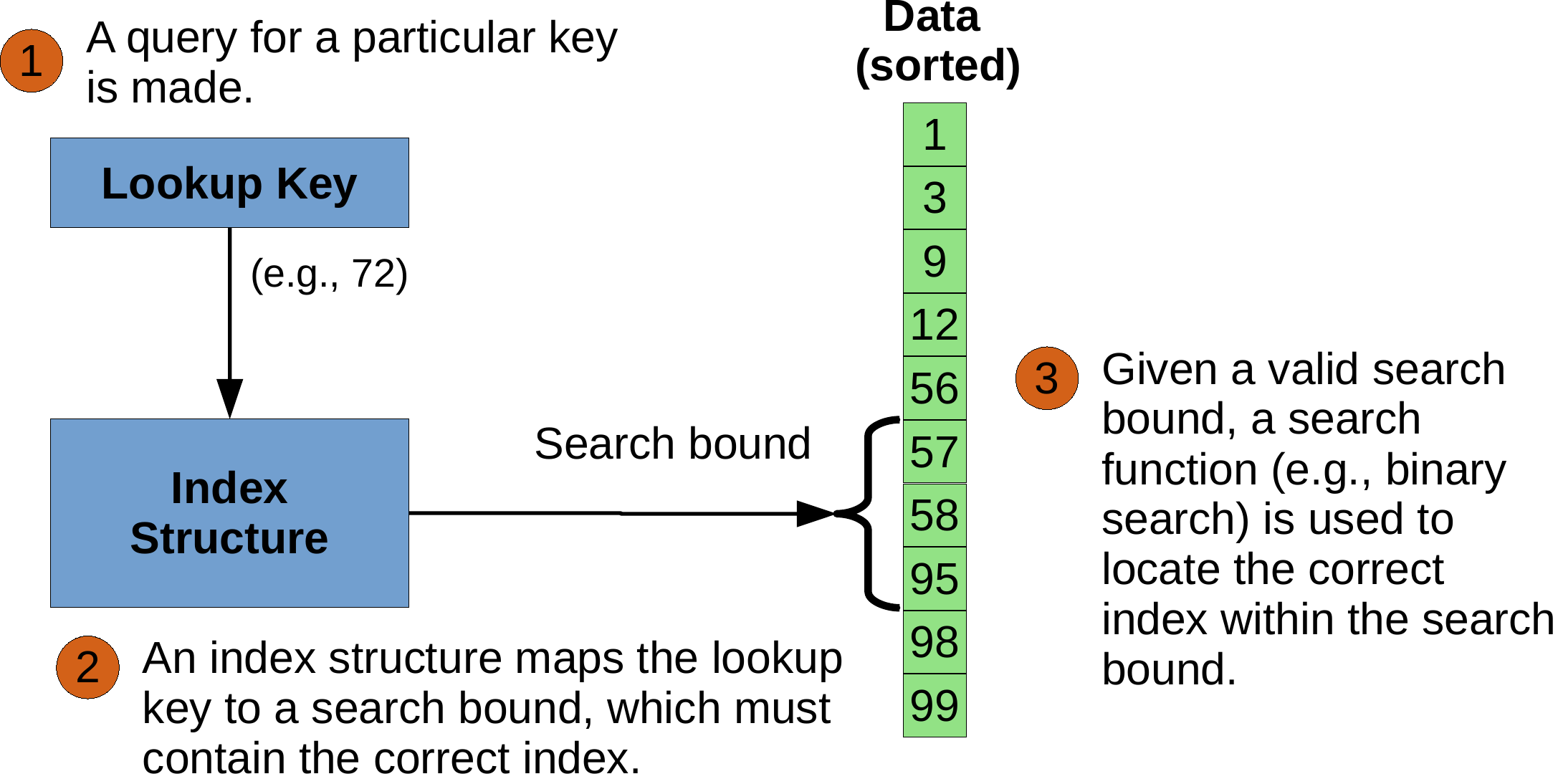}
  \caption{Index structures map each lookup key to a \emph{search bound}. This search bound must contain the ``lower bound'' of the key (i.e., the smallest key greater than or equal to the lookup key). The depicted search bound is valid for the lookup key 72 because the key 95 is in the bound. A search function, such as binary search, is used to locate the correct index within the search bound.}
  \label{fig:model}
\end{figure}

As depicted in Figure~\ref{fig:model}, we define an index structure $I$ over a zero-indexed sorted array $D$ as a mapping between an integer lookup key $x \in \mathbb{Z}$ and a \emph{search bound} $(lo, hi) \in (\mathbb{Z}^+ \times \mathbb{Z}^+)$, where $\mathbb{Z}^+$ is the positive integers and zero:

\begin{equation*}
I: \mathbb{Z} \to (\mathbb{Z}^+ \times \mathbb{Z}^+)
\end{equation*}

We do not consider indexes over unsorted data, nor do we consider non-integer keys. We assume that data is stored in a way supporting fast random access (e.g., an array).

Search bounds are indexes into $D$. A valid index structure maps any possible lookup key $x$ to a bound that contains the ``lower bound'' of $x$: the smallest key in $D$ that is greater than or equal to $x$. Formally, we define the lower bound of a key $x$, $LB(x)$, as:

\begin{equation*}
LB(x) = i \leftrightarrow \left[ D_i \geq x \land \lnot \exists j (j < i \land D_j \geq x) \right]
\end{equation*}


As a special case, we define the lower bound of any key greater than or equal to the largest key in $D$ as one more than the size of $D$: $LB(\max D) = |D|$. Our definition of ``lower bound'' corresponds to the C++ standard~\cite{url-cpp-lb}.

We say that an index structure is \emph{valid} if and only if it produces search bounds that contain the lower bound for every possible lookup key.

\begin{equation*}
\forall x \in \mathbb{Z} \left[ I(x) = (lo, hi) \to D_{lo} \leq LB(x) \leq D_{hi} \right]
\end{equation*}

Intuitively, this view of index structures corresponds to an \emph{approximate index}, an index that returns a search range instead of the exact position of a key. We are not the first to note that both traditional structures like B-Trees and learned index structures can be viewed in this way~\cite{ml_index, linear_reg_lookup}.

Given a valid index, the actual index of the lower bound for a lookup key is located via a ``last mile'' search (e.g., binary search). This last mile search only needs to examine the keys within the provided search bound (e.g., Figure~\ref{fig:model}). 

\subsection{Approximating the CDF}
\label{sec:cdf}
\begin{figure}
  \includegraphics[width=0.9\linewidth]{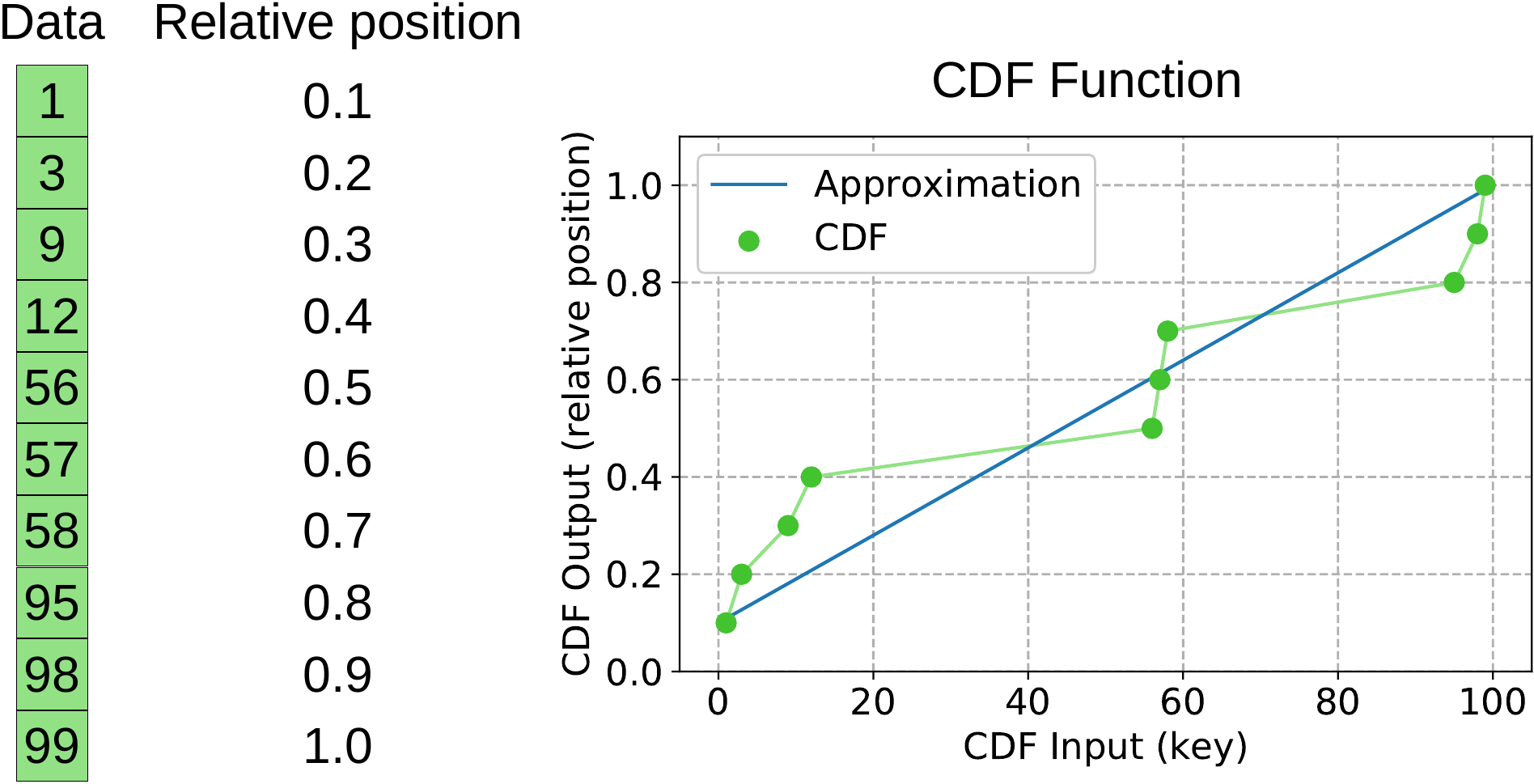}
  \caption{The cumulative distribution function (CDF) view of a sorted array.}
  \label{fig:cdf}
\end{figure}

Learned index structures use machine learning techniques ranging from deep neural networks to simple regression in order to model the \emph{cumulative distribution function}, or CDF, of a sorted array~\cite{ml_index}. Here, we use the term CDF to mean the function mapping keys to their relative position in an array. This is strongly connected to the traditional interpretation of the CDF from statistics: the CDF of a particular key $x$ is the proportion of keys less than $x$. Figure~\ref{fig:cdf} shows the CDF for some example data. 


Given the CDF of a dataset, finding the lower bound of a lookup key $x$ in a dataset $D$ with a CDF $CDF_D$ is trivial: one simply computes $CDF_D(x) \times |D|$. Learned index structures function by \emph{approximating} the CDF of the dataset using learned models (e.g., linear regressions). Of course, such learned models are never entirely accurate. For example, the blue line in Figure~\ref{fig:cdf} represents one possible imperfect approximation of the CDF. While imperfect, this approximation has a bounded error: the largest deviation from the blue line to the actual CDF occurs at key 12, which has a true CDF value of 0.4 but an approximated value of 0.24. The maximum error of this approximation is thus $0.4 - 0.24 = 0.16$ (some adjustments may be required for lookups of absent keys). Given this approximation function $A$ and the maximum error of $A$, we can define an index structure $I_A$ as such:

\begin{equation*}
I_A(x) = (A(x) - |D| \times 0.16, A(x) + |D| \times 0.16)
\end{equation*}

In other words, we can use the approximation of the CDF as an index structure by estimating the position of a given key and then computing the search bound of that estimate using the maximum error of the approximation. Note that this technique, while utilizing approximate machine learning techniques, \emph{never produces an incorrect search bound}.


One can view a B-Tree as a way of memorizing the CDF function for a given dataset: a B-Tree in which every $n$th key is inserted can be viewed as an approximate index with an error bound of $n-1$. At one extreme, if every key is inserted into the B-Tree, the B-Tree perfectly maps any possible lookup key to its position in the underlying data (an error bound of zero). Instead, one can insert every other key into a B-Tree in order to reduce the size of the index. This results in a B-Tree with an error bound of one: any location given by the B-Tree can be off by at most one position.


\section{Learned index structures}
\label{sec:lis}
In this work, we evaluate the performance of three different learned index structures: recursive model indexes (RMI), radix spline indexes (RS), and piecewise geometric model indexes (PGM).
We do not compare with a number of other learned index structures~\cite{alex,flood,fiting_tree} because tuned implementations could not be made publicly available.

While all three of these techniques approximate the CDF of the underlying data, the way these approximations are constructed vary. We next give a high-level overview of each technique, followed by a discussion of their differences.

\subsection{Recursive model indexes (RMI)}
\begin{figure}
  \centering
  \includegraphics[width=0.32\textwidth]{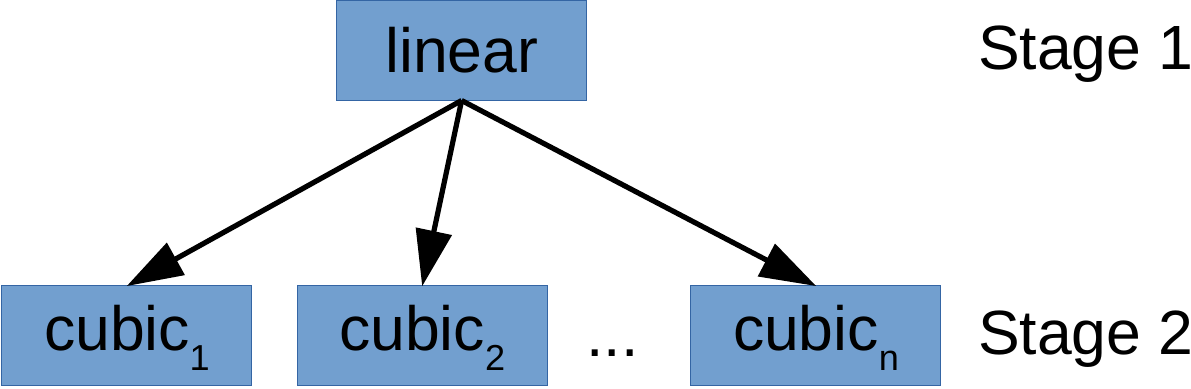}
  \caption{A recursive model index (RMI). The linear model (stage 1) makes a coarse-grained prediction. Based on this, one of the cubic models (stage 2) makes a refined prediction.}
  \label{fig:rmi}
\end{figure}

Originally presented by Kraska et al.~\cite{ml_index}, RMIs use a multi-stage model, combining simpler machine learning models together. For example, as depicted in Figure~\ref{fig:rmi}, an RMI with two stages, a linear stage and a cubic stage, would first use a linear model to make an initial prediction of the CDF for a specific key (stage 1). Then, based on that prediction, the RMI would select one of several cubic models to refine this initial prediction (stage 2).

\sparagraph{Structure} When all keys can fit in memory, RMIs with more than two stages are almost never required~\cite{cdfshop}. Thus, here we explain only two-stage RMIs for simplicity. See~\cite{ml_index} for a generalization to $n$ stages. A two-stage RMI is a CDF approximator $A$ trained on $|D|$ data points (key / index pairs). The RMI approximator $A$ is composed of a single first stage model $f_1$, and $B$ second-stage models $f^i_2$. The value $B$ is referred to as the ``branching factor'' of the RMI.

Formally, the RMI is defined as:

\begin{equation}
A(x) = f^{\left\lfloor B \times f_1(x) / |D| \right\rfloor}_2(x)
\label{eq:rmi}
\end{equation}

Intuitively, the RMI first uses the stage-one model $f_1(x)$ to compute a rough approximation of the CDF of the input key $x$. This coarse-grained approximation is then scaled between $0$ and the branching factor $B$, and this scaled value is used to select a model from the second stage, $f^i_2(x)$. The selected second-stage model is used to produce the final approximation. The stage-one model $f_1(x)$ can be thought of as partitioning the data into $B$ buckets, and each second-stage model $f_2^i(x)$ is responsible for approximating the CDF of only the keys that fall into the $i$th bucket.

Choosing the correct models for both stages ($f_1$ and $f_2$) and selecting the best branching factor for a particular dataset depends on the desired memory footprint of the RMI as well as the underlying data. In this work, we use the CDFShop~\cite{cdfshop} auto-tuner to determine these hyperparameters.

\sparagraph{Training} Let $(x, y) \in D$ be the set of key / index pairs in the underlying data. Then, an RMI is trained by adjusting the parameters contained in $f_1(x)$ and $f_2^i(x)$ to minimize:
\begin{equation}
\sum_{(x, y) \in D} \left( F(x) - y \right )^2
\label{eq:rmi_loss}
\end{equation}

Intuitively, minimizing Equation~\ref{eq:rmi_loss} is done by training ``top down'': first, the stage one model is trained, and then each stage 2 model is trained to fine-tune the prediction. Details can be found in~\cite{ml_index} and our implementation at~\cite{url-rmi}.

\subsection{Radix spline indexes (RS)}

\begin{figure}
    \centering
    \includegraphics[width=0.48 \textwidth]{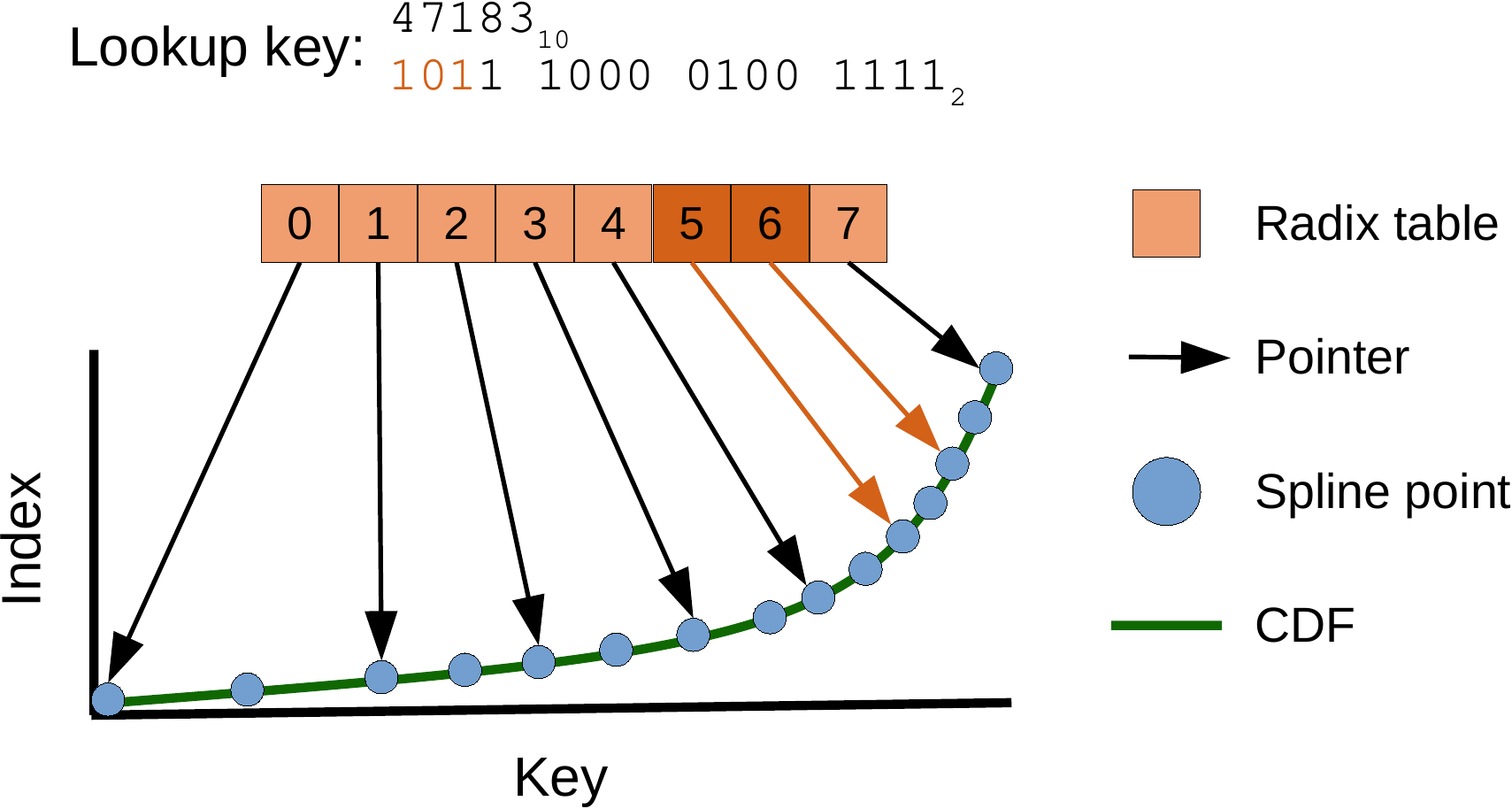}
    \caption{A radix spline index. A linear spline is used to approximate the CDF of the data. Prefixes of resulting spline points are indexed in a radix table to accelerate the search on the spline. Figure from~\cite{radix-spline}.}
    \label{fig:rs}
\end{figure}

An RS index~\cite{radix-spline} consists of a linear spline~\cite{spline} that approximates the CDF of the data and a radix table that indexes resulting spline points (cf., Figure~\ref{fig:rs}).
In contrast to RMI~\cite{ml_index}, and similar to FITing-Tree~\cite{fiting_tree} and PGM~\cite{pgm-index}, RS is built in a bottom-up fashion.
Uniquely, RS can be built in a single pass with a constant worst-case cost per element (PGM provides a constant amortized cost per element).


\sparagraph{Structure}
As depicted in Figure~\ref{fig:rs}, RS consists of a radix table and a set of spline points that define a linear spline over the CDF of the data.
The radix table indexes $r$-bit prefixes of the spline points and serves as an approximate index over the spline points.
Its purpose is to accelerate binary searches over the spline points.
The radix table is represented as an array containing $2^r$ offsets into the sorted array of spline points.
The spline points themselves are represented as key / index pairs.
To locate a key in a spline segment, linear interpolation between the two spline points is used.

Using the example in Figure~\ref{fig:rs}, a lookup in RS works as follows:
First, the $r$ most significant bits $b$ of the lookup key are extracted ($r=3$ and $b=$ {\color{orange} 101}).
Then, the extracted bits $b$ are used as an offset into the radix table to retrieve the offsets stored at the $b$th and the $b+1$th position (e.g., the 5th and the 6th position).
Next, RS performs a binary search between the two offsets on the sorted array of spline points to locate the two spline points that encompass the lookup key.
Once the relevant spline segment has been identified, it uses linear interpolation between the two spline points to estimate position of the lookup key in the underlying data.

\sparagraph{Training}
To build the spline layer, RS uses a one-pass spline fitting algorithm~\cite{spline} that is similar to the shrinking cone algorithm of FITing-Tree~\cite{fiting_tree}.
The spline algorithm guarantees a user-defined error bound.
At a high level, whenever the current error corridor exceeds the user-supplied bound, a new spline point is created.
Whenever the spline algorithm encounters a new $r$-bit prefix, a new entry is inserted into the pre-allocated radix table.

RS has only two hyperparameters (spline error and number of radix bits), which makes it straightforward to tune. In practice, few configurations need to be tested to reach a desired performance / size tradeoff on a given dataset~\cite{radix-spline}.

\subsection{Piecewise geometric model indexes (PGM)}
\begin{figure}
  \centering
  \includegraphics[width=0.4\textwidth]{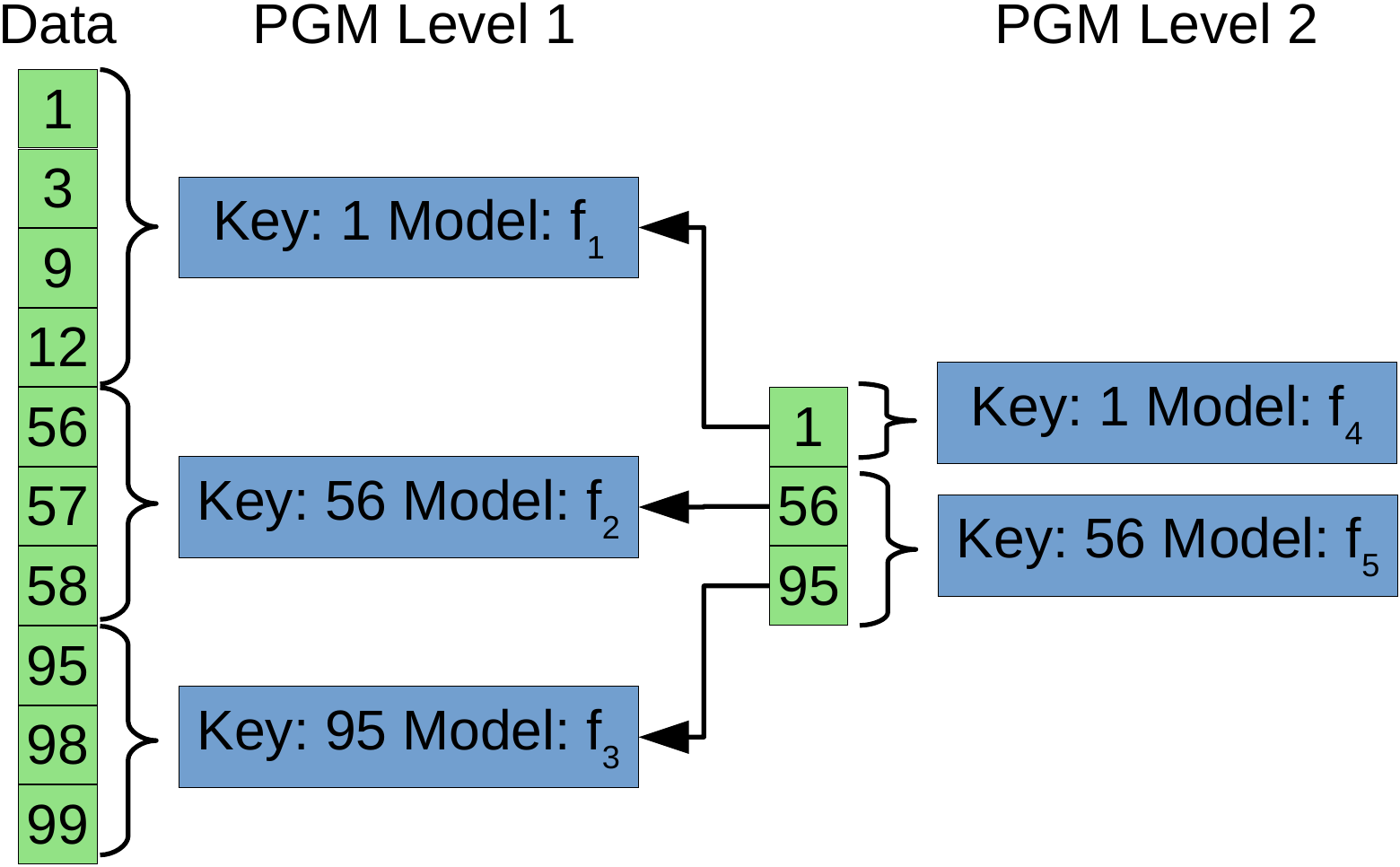}
  \caption{A piecewise geometric model (PGM) index.}
  \label{fig:pgm}
\end{figure}

The PGM index is a multi-level structure, where each level represents an error-bounded piecewise linear regression~\cite{pgm-index}. An example PGM index is depicted in Figure~\ref{fig:pgm}. In the first level, the data is partitioned into three segments, each represented by a simple linear model ($f_1, f_2, f_3$). By construction, each of these linear models predicts the CDF of keys in their corresponding segments to within a preset error bound. The partition boundaries of this first level are then treated as their own sorted dataset, and another error-bounded piecewise linear regression is computed. This is repeated until the top level of the PGM is sufficiently small.

\sparagraph{Structure} A piecewise linear regression partitions the data into $n+1$ segments with a set of points $p_0, p_1, \dots, p_n$. The entire piecewise linear regression is expressed as a piecewise function:

\begin{equation*}
  F(x) =
  \begin{cases}
    a_0 \times x + b_0 & \text{if $x < p_0$} \\
    a_1 \times x + b_1 & \text{if $x \geq p_0$ and $x < p_1$} \\
    a_2 \times x + b_2 & \text{if $x \geq p_1$ and $x < p_2$} \\
    \dots \\
    a_n \times x + b_n & \text{if $x \geq p_n$ and $x < p_n$} \\
  \end{cases}
\end{equation*}

Each regression in the PGM index is constructed with a fixed error bound $\epsilon$.
Such a regression can trivially be used as an approximate index. PGM indexes apply this trick recursively, first building an error-bounded piecewise regression model over the underlying data, then building another error-bounded piecewise regression model over the partitioning points of the first regression. Key lookups are performed by searching each index layer until the regression over the underlying data is reached.

\sparagraph{Training} Each regression is constructed optimally, in the sense that the fewest pieces are used to achieve a preset maximum error. This can be done quickly using the approach of~\cite{plr}. The first regression is performed on the underlying data, resulting in a set of split points (the boundaries of each piece of the regression) and regression coefficients. These split points are then treated as if they were a new dataset, and the process is repeated, resulting in fewer and fewer pieces at each level. Since each piecewise linear regression contains the fewest possible segments, the PGM index is optimal in the sense of piecewise linear models~\cite{pgm-index}.

Intuitively, PGM indexes are constructed ``bottom-up'': first, an error bound is chosen, and then a minimal piecewise linear model is found that achieves that error bound. This process is repeated until the piecewise models become smaller than some threshold. The PGM index can also handle inserts, and can be adapted to a particular query workload. We do not evaluate either capability here.

\subsection{Discussion}
\label{sec:lis_discussion}

RMIs, RS indexes, and PGM indexes all provide an approximation of the CDF of some underlying data using machine learning techniques. However, the specifics vary.

\sparagraph{Model types} While RS indexes and PGM indexes use only a single type of model (spline regression and piecewise linear regression, respectively), RMIs can use a wide variety of model types. This gives the RMI a greater degree of flexibility, but also increases the complexity of tuning the RMI. While both the PGM index and RS index can be tuned by adjusting just two knobs, automatically optimizing an RMI requires a more involved approach, such as~\cite{cdfshop}. Both the PGM index authors and the RS index authors mention integrating other model types as future work~\cite{radix-spline, pgm-index}.

\sparagraph{Top-down vs. bottom-up} RMIs are trained ``top down'', first fitting the topmost model and training subsequent layers to correct errors. PGM and RS indexes are trained ``bottom up'', first fitting the bottommost layer to a fixed accuracy and then building subsequent layers to quickly search the bottommost layer for the appropriate model. Because both PGM and RS indexes require searching this bottommost layer (PGM may require searching several intermediate layers), they may require more branches or cache misses than an RMI. While an RMI uses its topmost model to directly index into the next layer, avoiding a search entirely, the bottommost layer of the RMI does not have a fixed error bound; any bottom-layer model could have a large maximum error.

RS indexes and PGM indexes also differ in how the bottommost layer is searched. PGM indexes decompose the problem recursively, essentially building a second PGM index on top of the bottommost layer. Thus, a PGM index may have many layers, each of which must be searched (within a fixed range) during inference. On the other hand, an RS index uses a radix table to narrow the search range, but there is no guarantee on the search range's size. If the radix table provides a comparable search range as the upper level of a PGM index, then an RS index locates the proper final model with a comparatively cheaper operation (a bitshift and an array lookup). If the radix table does not provide a narrow search range, significant time may be spent searching for the appropriate bottom-layer model. 



\section{Experiments}
\label{sec:experiments}

Our experimental analysis is divided into six sections.

\begin{enumerate}[itemsep=0mm,itemindent=0mm,leftmargin=5mm]
\item{Setup (Section~\ref{sec:setup}): we describe the index structures, baselines, and datasets used.}
\item{Pareto analysis (Section~\ref{sec:pareto}): we analyze the size and performance tradeoffs offered by each index structure, including variations in dataset and key size. We find that learned index structures offer competitive performance.}
\item{Explanatory analysis (Section~\ref{sec:metrics}): we analyze indexes via performance counters (e.g., cache misses) and other descriptive statistics. We find that no single metric can fully account for the performance of learned structures.}
\item{CPU interactions (Section~\ref{sec:cpu}): we analyze how CPU cache and operator reordering impacts the performance of index structures. We find that learned index structures benefit disproportionately from these effects.}
\item{Multithreading (Section~\ref{sec:multithreading}): we analyze the throughput of each index in a multithreaded environment. We find that learned structures have comparatively high throughput, possibly attributable to the fact that they incur fewer cache misses per lookup.}
\item{Build times (Section~\ref{sec:build}): we analyze the time to build each index structure. We find that RMIs are slow to build compared to PGM and RS indexes, but that (unsurprisingly) no learned structure yet provides builds as fast as insert-optimized traditional index structures.}
\end{enumerate}

\subsection{Setup}
\label{sec:setup}
Experiments are conducted on a machine with 256 GB of RAM and an Intel(R) Xeon(R) Gold 6230 CPU @ 2.10GHz.

\subsubsection{Indexes}
\begin{table}
  \small
  \centering
  \begin{tabular}{lrrr}
    \toprule
    Method                           &  Updates & Ordered &   Type     \\
    \midrule                          
    PGM~\cite{pgm-index}             &  Yes     & Yes     & Learned \\
    RS~\cite{radix-spline}           &  No      & Yes     & Learned \\
    RMI~\cite{ml_index}              &  No      & Yes     & Learned \\
    \midrule
    BTree~\cite{url-stxbtree}        &  Yes     & Yes     & Tree \\
    IBTree~\cite{ibtree}             &  Yes     & Yes     & Tree \\
    FAST~\cite{fast}                 &  No      & Yes     & Tree \\
    \midrule
    ART~\cite{art}                   &  Yes     & Yes     & Trie \\
    FST~\cite{surf}                  &  Yes     & Yes     & Trie \\
    \midrule
    Wormhole~\cite{wormhole}         &  Yes     & Yes     & Hybrid hash/trie\\
    CuckooMap~\cite{url-simd-cuckoo} &  Yes     & No      & Hash \\
    RobinHash~\cite{url-tsl-robin}   &  Yes     & No      & Hash  \\
    \midrule
    RBS                              &  No      & Yes     & Lookup table \\
    BS                               &  No      & Yes     & Binary search \\
    \bottomrule
  \end{tabular}
  \caption{Search techniques evaluated}
  \label{tbl:techniques}
\end{table}

In this section, we describe the index structures we evaluate, and how we tune their size/performance tradeoffs. Table~\ref{tbl:techniques} lists each technique and its capabilities.

\sparagraph{Learned indexes} We compare with RMIs, PGM indexes, and RadixSpline indexes (RS), each of which are described in Section~\ref{sec:lis}. We use implementations tuned by each structure's original authors. RMIs are tuned using CDFShop~\cite{cdfshop}, an automatic RMI optimizer.  RS and PGM are tuned by varying the error tolerance of the underlying models. 

\sparagraph{Tree structures} We compare with several tree-structured indexes: the STX B-Tree (BTree)~\cite{url-stxbtree}, an interpolating BTree (IBTree)~\cite{ibtree}, the Adaptive Radix Trie (ART)~\cite{art}, the Fast Architectural-Sensitive Tree (FAST)~\cite{fast}, Fast Succinct Trie (FST)~\cite{surf}, and Wormhole~\cite{wormhole}.

For each tree structure, we tune the size/performance tradeoff by inserting a subset of the data as described in Section~\ref{sec:cdf}. To build a tree of maximum size with perfect accuracy, we insert every key. To build a tree with a smaller size and decreased accuracy, we insert every other key. We note that this technique, while simple, may not be the ideal way to trade space for accuracy in each tree structure. Specifically, ART may admit a smarter method in which keys are retained or discarded based on the fill level of a node. We only evaluate the simple and universal technique of inserting fewer keys into each structure, and leave structure-specific optimizations to future work.

\sparagraph{Hashing} While most hash tables do not support range queries, hash tables are still an interesting point of comparison due to their unmatched lookup performance. Unordered hash tables cannot be shrunk using the same technique as we use for trees.\footnote{Wormhole~\cite{wormhole}, which we evaluate, represents a state-of-the-art ordered hashing approach.} Therefore, we only evaluate hash tables that contain every key. We evaluate a standard implementation of a Robinhood hash table (RobinHash)~\cite{url-tsl-robin} and a SIMD-optimized Cuckoo map (CuckooMap)~\cite{url-simd-cuckoo}.

\sparagraph{Baselines} We also include two naive baselines: binary search (BS), and a radix binary search (RBS). Radix binary search~\cite{sosd} stores only the radix table used by the learned RS approach. We vary the size of the radix table to achieve different size/performance tradeoffs. 

\subsubsection{Datasets}
\begin{figure}
  \centering
  \includegraphics[width=0.9\linewidth]{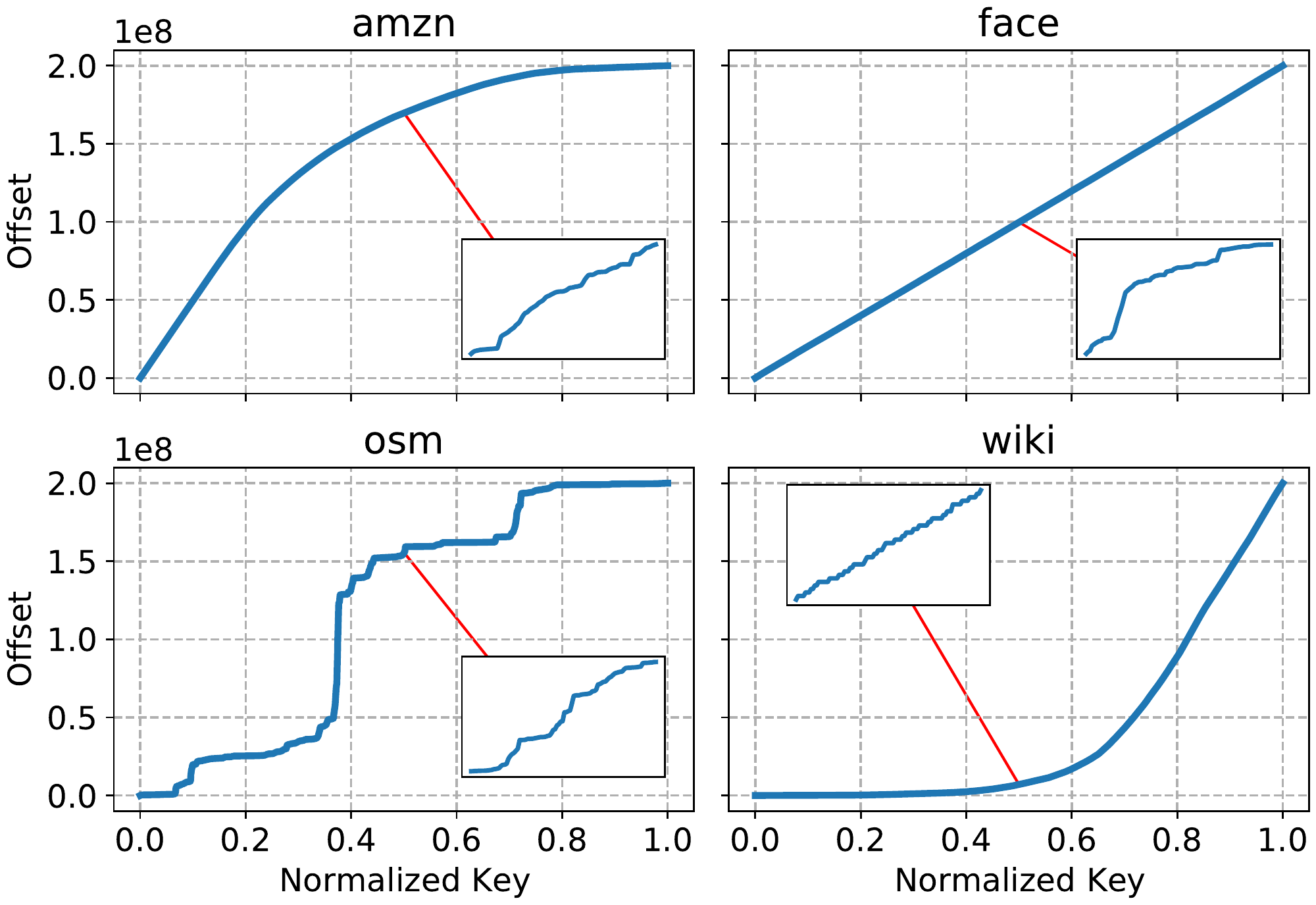}
  \caption{CDF plots of each testing dataset. The \fb dataset contains $\approx 100$ large outlier keys, not plotted.}
  \label{fig:cdfs}
\end{figure}

We use four real-world datasets for our evaluation. Each dataset consists of 200 million unsigned 64-bit integer keys. We test larger datasets in Section~\ref{sec:large}, and we test 32-bit datasets in Section~\ref{sec:bit32}. We generate 8-byte (random) payloads for each key. For each lookup, we compute the sum of these values to ensure the results are accurate.

\begin{itemize}[itemsep=0mm,itemindent=0mm,leftmargin=5mm,topsep=0pt]
\item{\textbf{amzn}: book popularity data from Amazon. Each key represents the popularity of a particular book.}
\item{\textbf{face}: randomly sampled Facebook user IDs~\cite{siptip}. Each key uniquely identifies a user.}
\item{\textbf{osm}: cell IDs from Open Street Map. Each key represents an embedded location.}
\item{\textbf{wiki}: timestamps of edits from Wikipedia. Each key represents the time an edit was committed.}
\end{itemize}

The CDFs of each of these datasets are plotted in Figure~\ref{fig:cdfs}. The zoom window on each plot shows 100 keys. While the ``zoomed out'' plots appear smooth, each CDF function is much more complex, containing both structure and noise.

For each dataset, we generate 10M random lookup keys. Indexes are required to return search bounds that contain the lower bound of each lookup key (see Section~\ref{sec:formulation}).

\sparagraph[?]{Why not test synthetic datasets} Synthetic datasets are often used to benchmark index structures, learned or otherwise~\cite{ml_index,pgm-index,art}. However, synthetic datasets are problematic for evaluating learned index structures. Synthetic datasets are either (1) entirely random, in which case there is no possibility of learning an effective model of the underlying data (although a model may be able to overfit to the noise),
or (2) drawn from a known distribution, in which case learning the distribution is trivial. Here, we focus only on datasets drawn from real world distributions, which we believe are the most important. For readers specifically interested in synthetic datasets, we refer to~\cite{sosd}.

\subsection{Pareto analysis}
\label{sec:pareto}
\begin{figure*}
  \centering
  \includegraphics[width=0.9\textwidth]{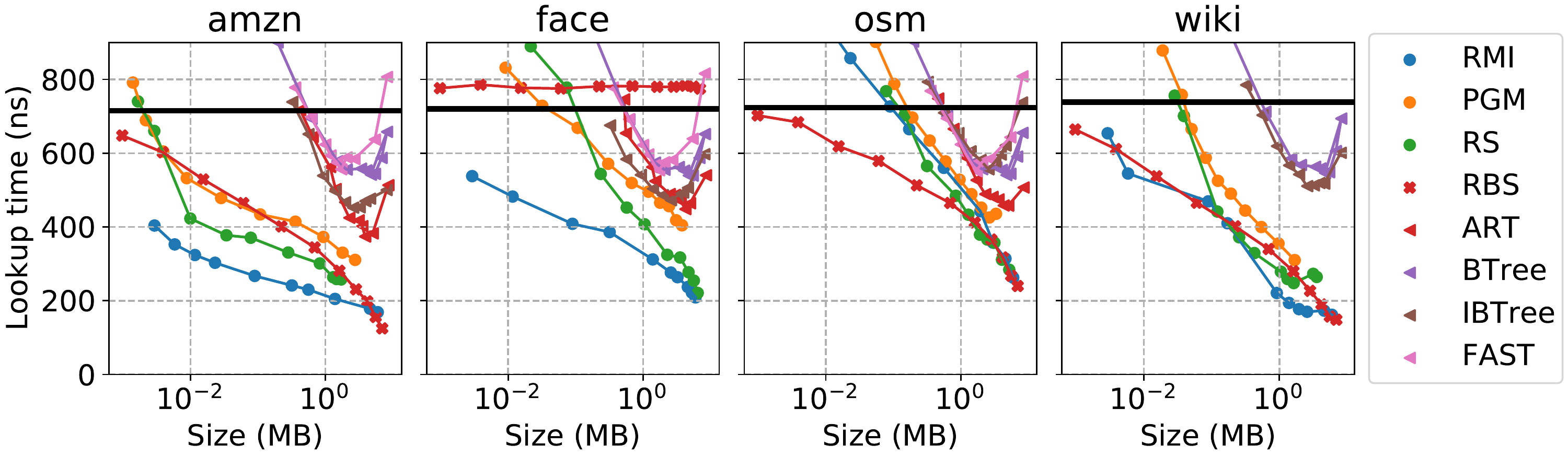}
  \caption{Performance and size tradeoffs provided by several index structures for four different datasets. The black horizontal line represents the performance of binary search (which has a size of zero). Extended plots with all techniques are available here: \url{https://rm.cab/lis1}}
  \label{fig:pareto}
\end{figure*}

A primary concern of index structures is lookup performance: given a query, how quickly can the correct record be fetched? However, size is also important: with no limits, one could simply store a  lookup table and retrieve the correct record with only a single cache miss. Such a lookup table would be prohibitively large in many cases, such as 64-bit keys. Thus, we consider \emph{the performance / size tradeoff} provided by each index structure, plotted in Figure~\ref{fig:pareto}.

For each index structure, we selected ten configurations ranging from minimum to maximum size. While different applications may weigh performance and size differently, all applications almost surely desire a \emph{Pareto optimal} index: an index for which no alternative has both a smaller size and improved performance. For the \books and \wiki datasets, learned structures are Pareto optimal up to a size of 100MB, at which point the RBS lookup table becomes effective. For \face, learned structures are Pareto optimal throughout.

\sparagraph{Poor performance on \osm} Both traditional and learned index structures fail to outperform RBS  on the \osm dataset for nearly any size. The poor performance of learned index structures can be attributed to the \osm's dataset lack of local structure: even small pieces of the CDF exhibit difficult-to-model erratic behavior. This is an artifact of the technique used to project the Earth into one-dimensional space (a Hilbert curve).  In Section~\ref{sec:metrics}, we confirm this intuition by analyzing the errors of the learned models; all three learned structures required significantly more storage to achieve errors comparable to those observed on the other datasets. Simply put, learned structures perform poorly on \osm because \osm is difficult to learn. Because \osm is a one-dimensional projection of multi-dimensional data, a multi-dimensional learned index~\cite{flood} may yield improvements.


\sparagraph{Performance of PGM} In~\cite{pgm-index}, the authors showed that ``the PGM-index dominates RMI,'' contradicting our previous experience that the time spent on searches between the layers of the index outweighed the benefits of having a lower error. Indeed, in our experimental evaluation we found that the PGM index performs significantly worse than RMI on 3 out of the 4 datasets and slightly worse on \osm.
After contacting the authors of~\cite{pgm-index}, we found that their RMI implementation was missing several key optimizations: their RMI only used linear models rather than tuning different type of models as proposed in~\cite{ml_index,cdfshop}, and omitted some optimizations for RMIs with only linear models.\footnote{\small We shared our RMI implementation with Ferragina and Vinciguerra before the publication of~\cite{pgm-index}, but since~\cite{pgm-index} was already undergoing revision, they elected to continue with their own RMI implementation instead, without note. All PGM results in this paper are based on Ferragina and Vinciguerra's tuned PGM code as of May 18th, 2020.} This highlights how implementation details can affect experimental results, and the importance of having a common benchmark with strong implementations. We stress that our results are the first to compare RMI and PGM implementations tuned by their respective authors.

\sparagraph{Performance of RBS} RBS exhibits substantially degraded performance on \face compared to other datasets. This is due to a small number ($\approx 100$) of outliers in the \face dataset: most keys fall within $(0, 2^{50})$, but the outliers fall in $(2^{59},2^{64}-1)$. These outliers cause the first 16 prefix bits of the RBS lookup table to be nearly useless.
One could adjust RBS to handle this simple case (when all outliers are at one end of the dataset), but in general such large jumps in values represents a severe weakness of RBS. ART~\cite{art} can be viewed as a generalization of RBS to handle this type of skew regardless of where it occurs in the dataset.

On other datasets, RBS is surprisingly competitive, often outperforming other indexes. This is partially explained by the low inference time required by RBS: getting a search bound requires only a bit shift and an array lookup. When the prefixes of keys are distributed uniformly across a range, an RBS with a radix table of size $2^b$ provides equally accurate bounds as a binary search tree with $b$ levels, but requires only a single cache miss. When the keys are heavily skewed (as is the case with \face), the radix table is nearly useless.


\sparagraph{Tree structures are non-monotonic} All tree structures tested (ART, BTree, IBTree, and FAST) become less effective after a certain size. For example, the largest ART index for the \books data occupies nearly 1GB of space, but has worse lookup performance than an ART index occupying only 100MB of space. This is because, at a certain point, performing a binary search on a small densely-packed array becomes more efficient than traversing a tree. As a result, tree structures show non-monotonic behavior in Figure~\ref{fig:pareto}. 

\sparagraph[?]{Indexes slower than binary search} At extremely small or large sizes, some index structures perform worse than binary search. In both cases, this is because some index structures are unable to provide sufficiently small search bounds to make up for the inference time required. For example, on the \osm dataset, very small RMIs barely narrow down the search range at all. Because this small RMIs fit is so poor (analyzed later, Figure~\ref{fig:metrics}), the time required to execute the RMI model and produce the search bound is comparatively worse than executing a binary search on the entire dataset.

\begin{figure}[t]
  \includegraphics[width=\linewidth]{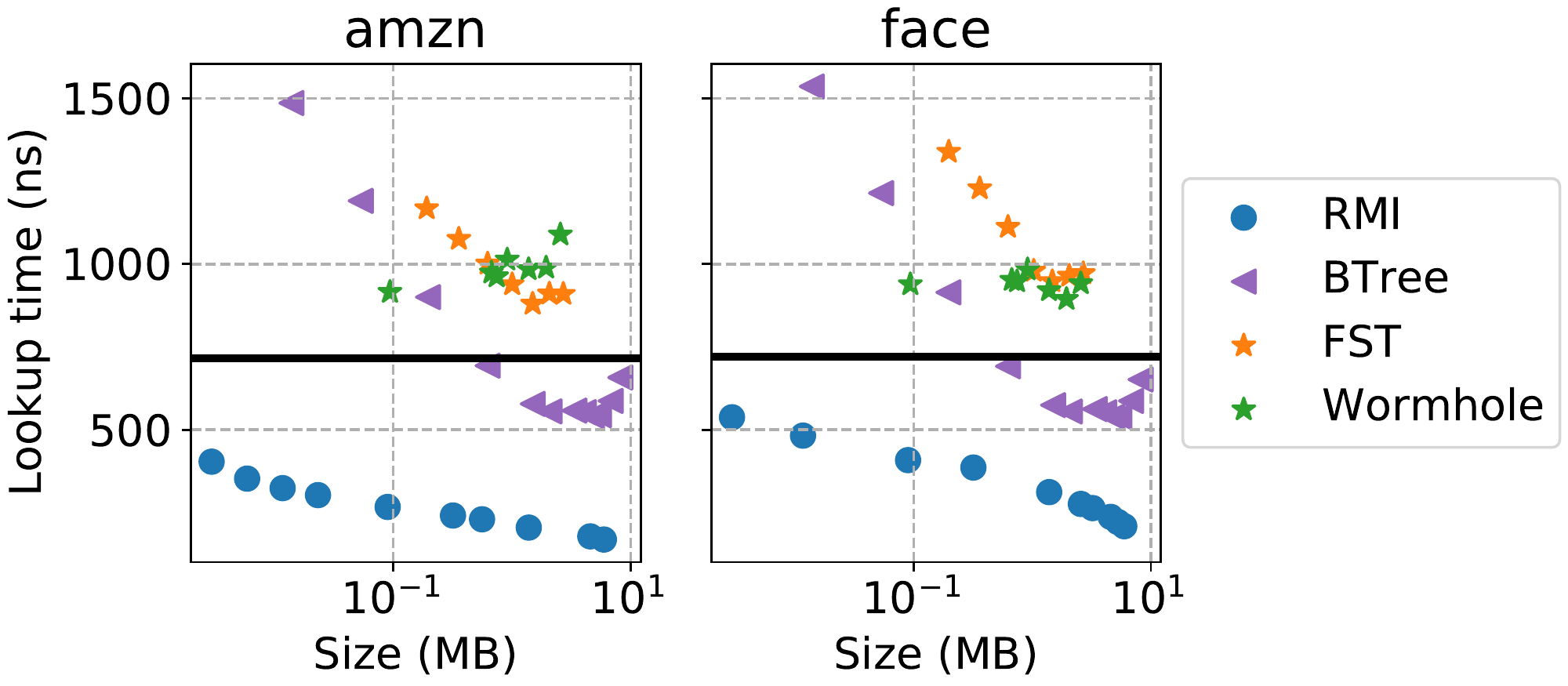}
  \caption{Performance of index structures built for strings (stars) on our integer datasets.}
  \label{fig:pareto_strs}
\end{figure}

\sparagraph{Structures for strings} Many recent works on index structures have focused on indexing keys of arbitrary length (e.g., strings)~\cite{surf, wormhole}. 
For completeness, we evaluated two structures designed for string keys -- FST and Wormhole -- in Figure~\ref{fig:pareto_strs}. Unsurprisingly, neither performed as well as binary search. These string indexes contain optimizations that assume that comparing two keys is expensive. These optimizations translate to overhead when considering only integer keys, which can be compared in a single instruction. ART, an index designed for both string and integer data, does so by indexing one key-byte per radix tree level.

\begin{table}
  \centering
  \small
  \begin{tabular}{lrr}
\toprule
    Method &      Time &      Size \\
\midrule
       PGM & 326.48 ns &   14.0 MB \\
        RS & 266.58 ns &    4.0 MB \\
       RMI & 180.90 ns &   48.0 MB \\
     BTree & 482.11 ns &  166.0 MB \\
    IBTree & 446.55 ns &    9.0 MB \\
      FAST & 435.33 ns &  102.0 MB \\
        BS & 741.69 ns &    0.0 MB \\
\midrule
 CuckooMap & 114.50 ns & 1541.0 MB \\
 RobinHash &  93.69 ns & 6144.0 MB \\
\bottomrule
\end{tabular}

  \caption{The fastest variant of each index structure compared against two hashing techniques on the \books dataset.}
  \label{tbl:hashing}
\end{table}

\sparagraph{Hashing} Hashing provides $O(1)$ time point lookups. However, hashing differs from both traditional and learned indexes in a number of ways: first, hashing generally does not support lower bound lookups.\footnote{\small Wormhole, evaluated in Figure~\ref{fig:pareto_strs}, is a hash-based technique that provides ordering, but is primarily optimized for strings.} Second, hash tables generally have a large footprint, as they store every key.
We evaluate two hashing techniques -- a Cuckoo hash table~\cite{url-simd-cuckoo} and a Robinhood hash table~\cite{url-tsl-robin}. 
We found that a load factor of 0.99 and 0.25 (respectively) maximized lookup performance.

Table~\ref{tbl:hashing} lists the size and lookup performance of the best-performing (and thus often largest) variant of each index structure and both hashing techniques for a 32-bit version\footnote{\small The SIMD Cuckoo implementation only supports 32-bit keys.} of the \books dataset (results similar for others). Unsurprisingly, both hashing techniques offer superior point-lookup latency compared to traditional and learned index structures. This decreased latency comes at the cost of a larger in-memory footprint. For example, CuckooMap provides a 114ns lookup time compared to the 180ns provided by the RMI, but CuckooMap uses over 1GB of memory, whereas the RMI uses only 48MB. When range lookups and memory footprint are not concerns, hashing is a clear choice.



\subsubsection{Larger datasets}
\label{sec:large}
\begin{figure*}[ht]
  \centering
  \includegraphics[width=0.9\textwidth]{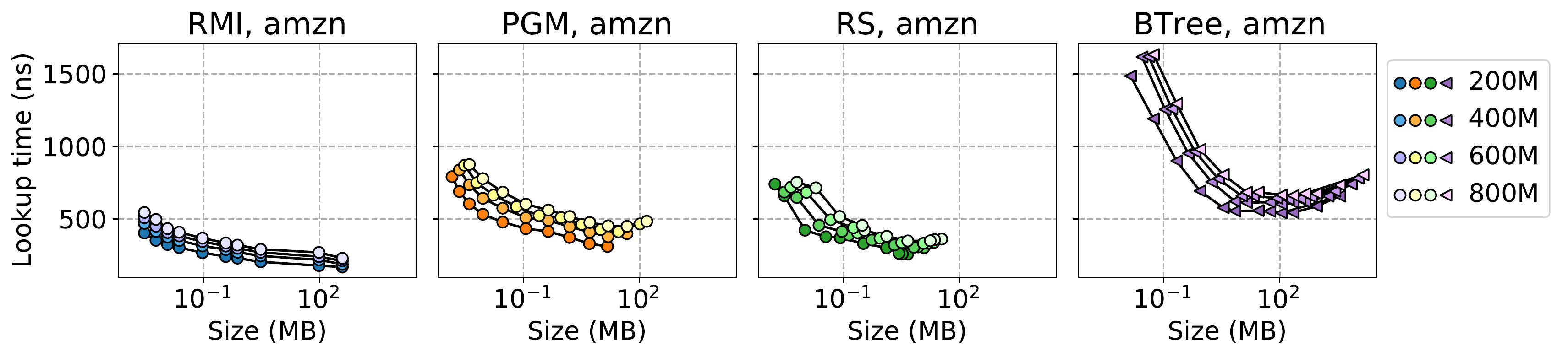}
  \caption{Performance / size tradeoffs for datasets of various sizes (200M, 400M, 600M, and 800M keys) for the \books dataset. The \fb and \wiki datasets were not sufficiently large to compare. Extended plots with all techniques and the \osm dataset are available here: \url{https://rm.cab/lis2}}
  \label{fig:pareto_sz}
\end{figure*}

Figure~\ref{fig:pareto_sz} shows the performance / size tradeoff for each learned structure and a BTree for four different data sizes of the \books dataset, ranging from 200M to 800M.
All three learned structures are capable of scaling to larger dataset sizes, with only a logarithmic slowdown (as is expected from the final binary search step). For example, consider an RMI that produces an average search bound that spans 128 keys. Such a bound requires 7 steps of binary search. If the dataset size doubles, an RMI of equal size is likely to return bounds that are twice as large: one could expect an RMI of equal size to produce search bounds that span 256 keys. Such a bound requires only 8 total (1 additional) binary search steps. Thus, learned index structures scale to larger datasets in much the same way as BTrees. If larger datasets have more pronounced and modelable patterns, learned index structures may provide better scaling.


\subsubsection{32-bit datasets}
\label{sec:bit32}
\begin{figure*}
  \includegraphics[width=\textwidth]{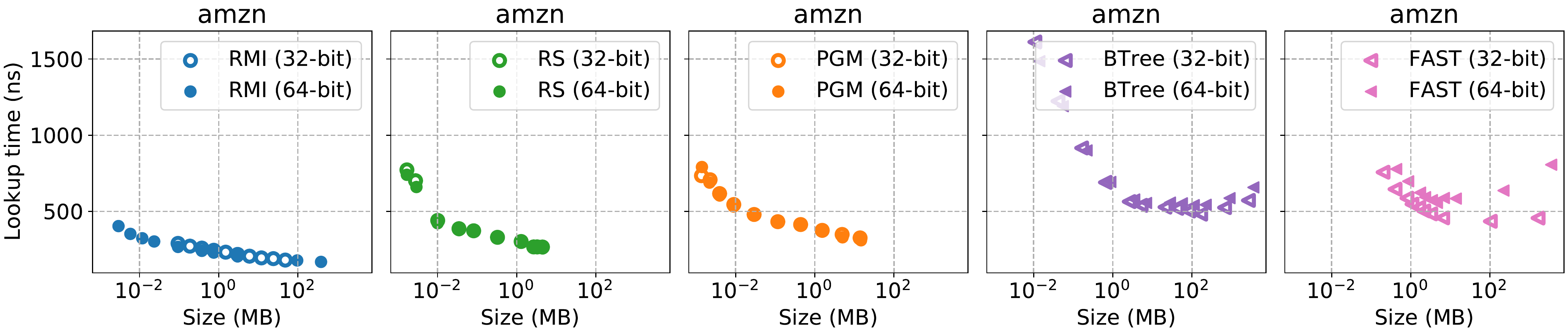}
  \caption{Performance / size tradeoff for 32 and 64 bit keys. While decreasing the key size to 32-bits has a minimal impact on learned structures, the ability to pack more values into a single cache line improve the performance of tree structures.}
  \label{fig:32_vs_64}
\end{figure*}

Other sections evaluate 64-bit datasets. Here, we scale down the \books dataset from 64 to 32 bits, and compare the performance of the three learned index structures, BTrees, and FAST. The results are plotted in Figure~\ref{fig:32_vs_64}.

For learned structures, the performance on 32-bit data is nearly identical to performance on 64-bit data. Our implementations of RS and RMI both transform query keys to 64-bit floats, so this is not surprising. We attempted to perform computations on 32-bit keys using 32-bit floats, but found that the decreased precision caused floating point errors.
The PGM implementation uses 32-bit computations for 32-bit inputs, achieving some modest performance gains.

For both tree structures, the switch from 64-bit to 32-bit keys allows twice as many keys to fit into a single cache line, improving performance. For FAST, which makes heavy use of AVX-512 streaming operations, doubling the number of keys per cache line essentially doubles computational throughput as well, as each operator can work on 16 32-bit values simultaneously (as opposed to 8 64-bit values). 


\subsubsection{Search function}
\begin{figure}[ht]
  \includegraphics[width=\linewidth]{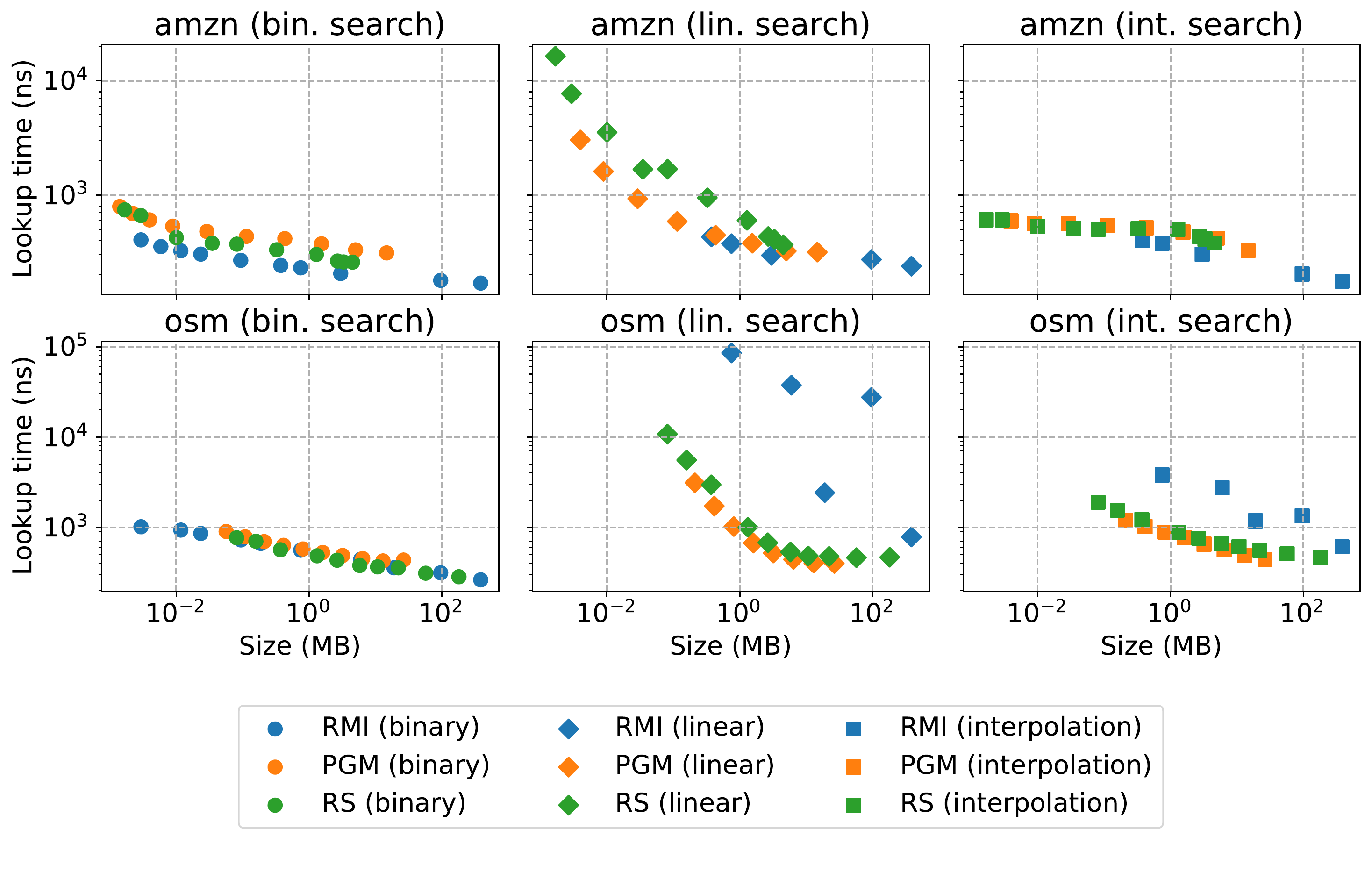}
  \caption{A comparison of ``last mile'' (Section~\ref{sec:formulation}) search techniques for the \osm and \books datasets.}
  \label{fig:linear_vs_binary}
\end{figure}

Normally, we use binary search to locate the correct key within the search bound provided by the index structure. However, other search techniques can be used. Figure~\ref{fig:linear_vs_binary} evaluates binary, linear, and interpolation search for each learned structure and the RBS baseline on \osm and \books.


We observed that binary search (first column) was always faster than linear search (second column). This aligns with prior work that showed binary search being effective until the data size dropped below a very small threshold~\cite{opt_binary_search}.

Interpolation search (third column) behaves similarly to binary search on the \books dataset, even offering improved performance on average ($\approx 2\%$). This was surprising, because interpolation search works by assuming that keys are uniformly distributed between two end points. If this were the case, one would expect a learned index to learn this distribution, subsuming any gains from interpolation search. However, because the learned structures have a limited size, there can be many segments of the underlying data that exhibit linear behavior that the learned structure does have the capacity to learn. For the \osm dataset, which is relatively complex, interpolation search does not provide a benefit, and is often slower than binary search. This is unsurprising, since interpolation search works best on smooth datasets.

One could also integrate more complex interpolation search techniques, such as SIP~\cite{siptip}. One difficulty with incorporating SIP is the precomputation steps, which vary depending on the search bound used. Integrating an exponential search~\cite{exponential-search} technique could also be of interest, although it is not immediately clear how to integrate a search bound. We leave such investigations to future work.



\subsection{Explaining the performance}
\label{sec:metrics}

In this section, we investigate \emph{why} learned index structures have such strong performance and size properties. While prior work~\cite{ml_index} attributed this to decreased branching and instruction count, we discovered that the whole story was more complex. None of model accuracy, model size (or ``precision gain'', the combination of the two in~\cite{ml_index}), cache misses, instruction count, or branch misses can fully account for learned index structures' performance. 

\begin{figure*}[ht]
  \includegraphics[width=\textwidth]{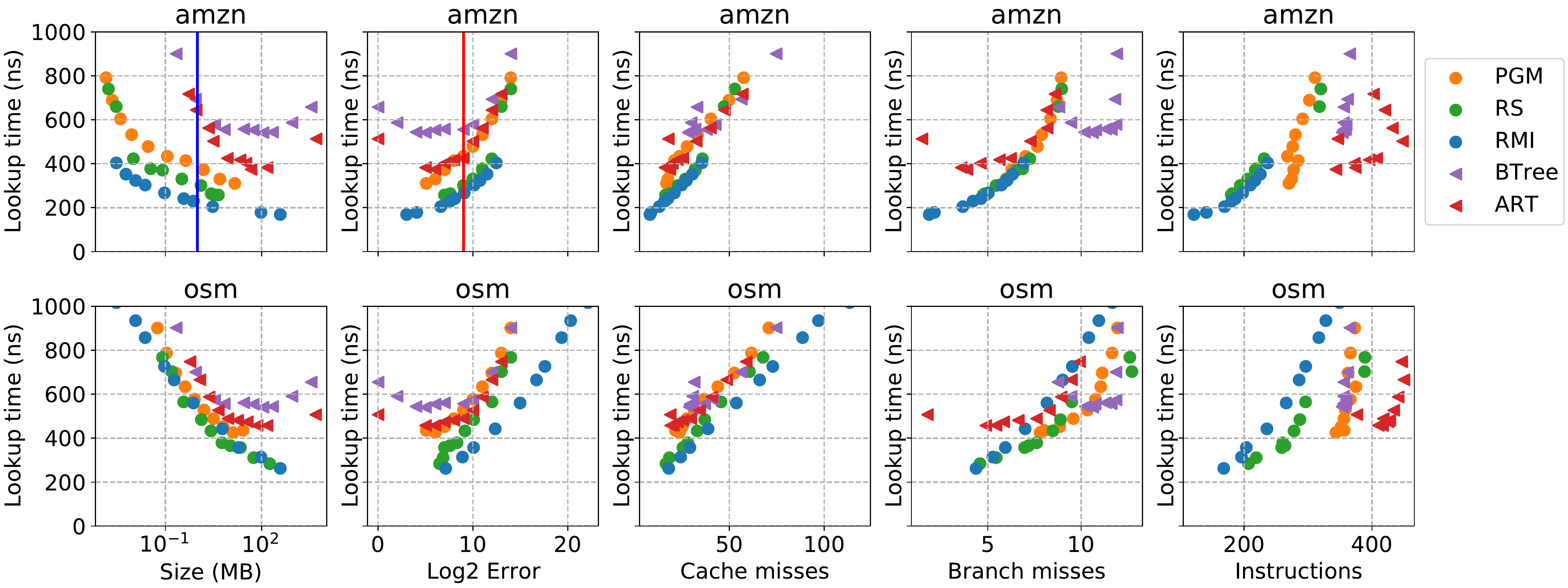}
  \caption{Various metrics compared with lookup times across index structures and datasets. No single metric can fully explain the performance of different index structures, suggesting a multi-metric analysis is required. Extended plots for all techniques and datasets are available here: \url{https://rm.cab/lis5}}
  \label{fig:metrics}
\end{figure*}

Figure~\ref{fig:metrics} shows the correlation between lookup time and various performance characteristics of learned index structures, BTrees, and ART for the \books and \osm datasets. The first column shows the total in-memory size of each model, the second column shows average $\log_2$ search bound size (i.e., the expected number of binary search steps required), the third column shows last-level cache misses, the fourth column shows branch mispredictions, and the fifth column shows instruction counts. One can visually dismiss any single metric as explanatory: any vertical line corresponds to structures that are equal on the given metric, but exhibit different lookup times. For example, at a size of 1MB, RMIs achieve a latency of 220ns on \books, but a BTree with the same size achieves a latency of 650ns (blue vertical line).

The second column (``$\log_2$ error''), is especially interesting. Learned indexes must balance inference time with model error~\cite{cdfshop}. For example, with a $\log_2$ error of 7, an RMI achieves a lookup time of 250ns on the \books dataset, but the PGM index with the same $\log_2$ error achieves a latency of 480ns (red vertical line). In other words, even though the average size of the search bound generated by both structures was the same, the RMI still achieved faster lookup times. This is attributable to the higher inference time of the PGM index. Of course, other factors, such as overall model size, must be taken into account as well.



\sparagraph{Analysis} In order to statistically test each potential explanatory factor, we performed a linear regression analysis using every index structure on all four datasets at 200 million 64-bit keys. The results indicated that \emph{cache misses, branch misses, and instruction count had a statistically significant effect on lookup time ($p < 0.001$)}, whereas size and $\log_2$ error did not ($p > 0.15$). To be clear, this means that \emph{given the branch misses, cache misses, and instruction counts}, the size and $\log_2$ error do not significantly affect performance. This does not mean that the $\log_2$  error and size do not have an impact on cache misses; just that the relevant variation in lookup time explained by model size and $\log_2$ error is accounted for fully in the other measures. 

Overall, a regression on cache misses, branch misses, and instruction count explained 95\% of the variance ($R^2 = 0.955$). This means that 95\% of the variation we observed in our experiments can be explained by a linear relationship between cache misses, branch misses, instructions, and lookup latency.
The standardized regression coefficients for cache misses, branch misses, and instruction misses were $0.85$, $-0.28$, and $0.50$, respectively. Standardized regression coefficients can be interpreted as the number of standard deviations that a particular measure needs to increase by, assuming the other measures stay fixed, in order to increase the output by one standard deviation; in other words, these coefficients are descriptive of the variations within our measurements, not of the actual hardware impact of the metrics (although these are obviously related).

\sparagraph{Interpretation: branch misses} While the magnitude of standardized regression coefficients are not useful on their own, their sign can provide interesting insights. Surprisingly, the coefficient on branch misses is negative. This does not mean that an increased number of branch misses leads to increased model performance. Instead, the negative coefficient means that \emph{for a fixed number of cache misses and instructions}, the tested indexes that incurred more branch misses performed better. In other words, indexes are getting significant value from branch misses; when an index incurs a branch miss, it does so in such a way that reduces lookup time more than an hypothetical alternative index that uses the same number of instructions and cache misses. 

We offer two possible explanations for this surprising observation. First, structures may be over-optimized to avoid branching, trading additional cache misses or instructions to reduce branching. Second, indexes that experience more branch misses may benefit from speculative loads on modern hardware. We leave further investigation to future work.

\sparagraph[?]{Interpretation: what metrics matter} If there is a single metric that explains the performance of learned index structures, we were unable to find it. Any of model size, $\log_2$ error, cache misses, branch misses, and instruction count alone are not enough to determine if one index structure will be faster than another. Linear regression analysis suggests that cache misses, branch misses, and instruction counts are all significant, and account for model size and $\log_2$ error. Of the significant measures, cache misses had the largest explanatory power. This is consistent with indexes being latency-bound (i.e., limited by the round-trip time to RAM).

The vast majority of cache misses for RMIs happen during the last-mile search. Two-layer RMIs require at most two cache misses for inference (potentially only one if the RMI's top layer is small enough). 
On the other hand, for a full BTree, no cache misses happen during the final search at all, but BTrees generally require at least one cache miss per level of the tree. 
Cache misses also help explain performance differences between RMI and PGM: since each additional PGM layer likely requires a cache miss at inference time, a large RMI with low $\log_2$ error will incur fewer cache misses than a large PGM index with a similar $\log_2$ error (e.g., \books in Figure~\ref{fig:metrics}). When an RMI is not able to achieve a low $\log_2$ error, this advantage vanishes, as more cache misses are required during the last-mile search (e.g., \osm in Figure~\ref{fig:metrics}).

Current implementations of learned index structures seem to prioritize fast inference time over $\log_2$ error. This makes sense, since a linear increase in $\log_2$ error only leads to a logarithmic increase in lookup time (due to binary search). However, our analysis suggests that a learned index structure could use significantly more cache misses if it could accurately pinpoint the cache line containing the lookup key. We experimented with multi-stage RMIs ($> 10$ levels), but were unable to achieve such an accuracy. This could be an interesting direction for future work.

We encourage future development of index structures to take into account cache misses, branch misses, and instruction counts. Since all three of these metrics have a statistically significant impact on performance, ignoring one or two of them in favor of the other may lead to poor results. While we cannot suggest a single metric for evaluating index structures, if one must select a single metric, our analysis suggests that cache misses are the most significant.

\begin{figure}
  \centering
  \includegraphics[width=0.9\linewidth]{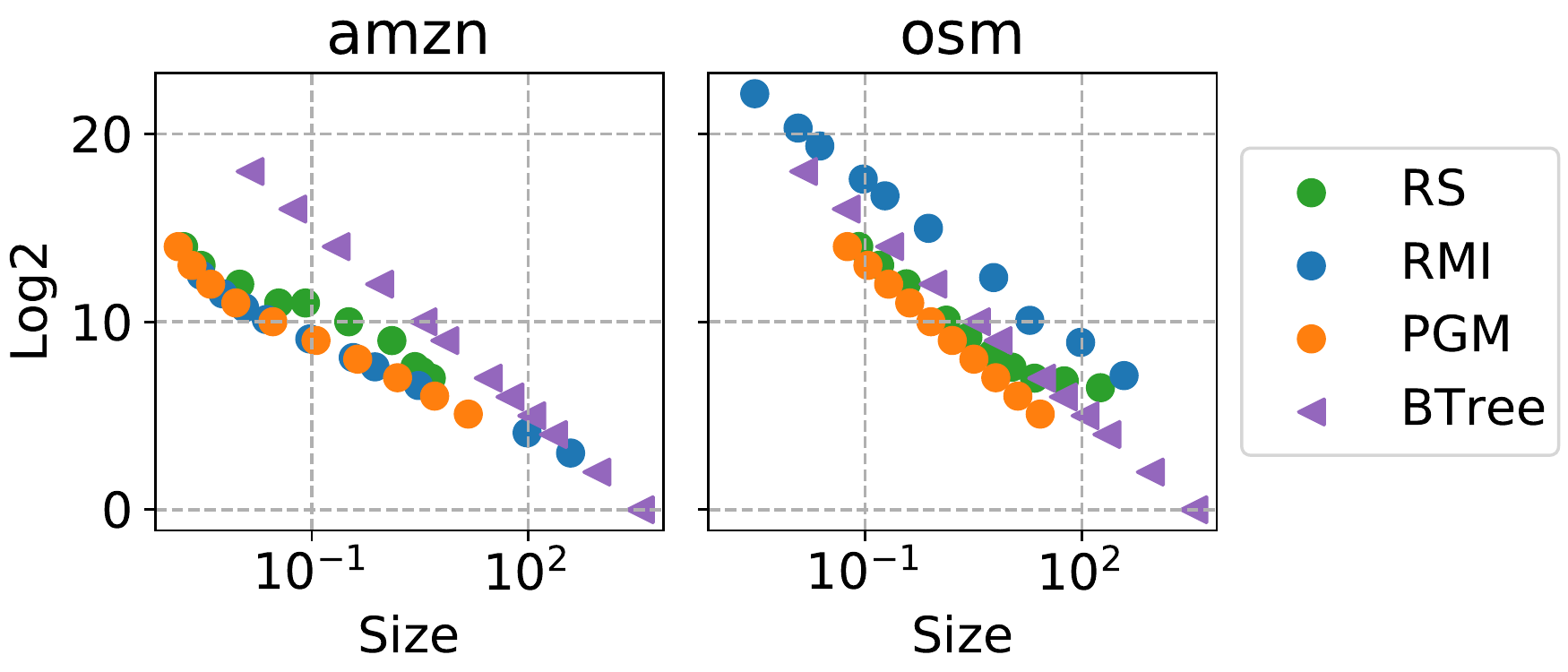}
  \caption{Size and $\log_2$ error bound of various index structures. When evaluated as a compression technique, learned index structures can be evaluated purely based on their size and $\log_2$ error. Extended plots are available here: \url{https://rm.cab/lis7}}
  \label{fig:size_vs_error}
\end{figure}

\sparagraph{Learned indexes as compression} A common view of learned index structures is to think of learned indexes as a lossy compression of the CDF function~\cite{pgm-index, ml_index}. In this view, the goal of a learned index is similar to lossy image compression (like JPG): come up with a representation that is smaller than the CDF with minimal information loss. The quality of a learned index can thus be judged by just two metrics: the size of the structure, and the $\log_2$ error (information loss).
Figure~\ref{fig:size_vs_error} plots these two metrics for the three learned index structures and BTrees. These plots indicate that the information theoretic view, while useful, is not fully predictive of index performance. For example, for \face, all three structures have very similar size and $\log_2$ errors after 1MB. However, some structures are substantially faster than others at a fixed size (Figure~\ref{fig:pareto}).

We encourage researchers and practitioners to familiarize themselves with the information theoretic view of learned index structures, but we caution against ending analysis at this stage. For example, an index structure that achieves optimal compression (i.e., an optimal size to $\log_2$ error ratio) is \emph{not necessarily} going to outperform an index with suboptimal compression. The simplest way this could occur is because of inference time: if the index structure with superior compression takes a long time to produce a search bound, an index structure that quickly generates less accurate search bounds may be superior. 
However, if one assumes that storage mediums are arbitrarily slow (i.e., search time is strictly dominated by the size of search bound), then there is merit in viewing learned index structures as a pure compression problem, and investigating more advanced compression techniques for these structures~\cite{pgm-index} could be fruitful.


\subsection{CPU interactions}
\label{sec:cpu}
Many prior works on both learned and non-learned index structures (including those by authors of this work) have evaluated their index structures by repeatedly performing lookups in a tight loop. While convenient and applicable to many applications, this experimental setup may exaggerate the performance of some index structures due, in part, to \emph{caching} and \emph{operator reordering}. 

\subsubsection{Caching}
\label{sec:cache}

\begin{figure*}
  \includegraphics[width=\textwidth]{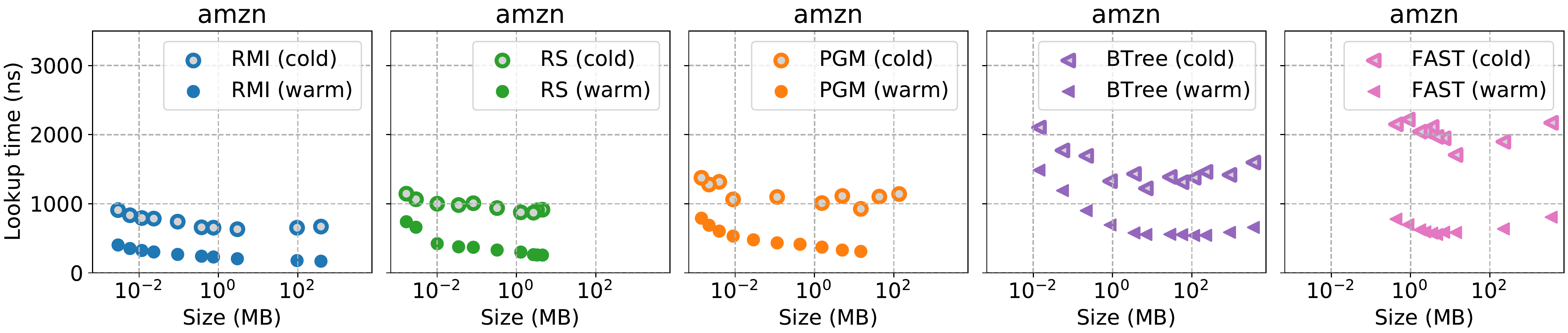}
  \caption{The performance impact of having a cold cache for various index structures. Extended plots with all techniques are available here: \url{https://rm.cab/lis3}}
  \label{fig:cache}
\end{figure*}

Executing index lookups in a tight loop, as it is often done to evaluate an index structure, will cause nearly all of the CPU cache to be filled with the index structure and underlying data. Since accessing a cached value is significantly faster (10s of nanoseconds) than accessing an uncached value ($\approx 100$ nanoseconds), this may cause such tight-loop experiments to exaggerate the performance of an index structure.

The amount of data that will remain cached from one index lookup to another is clearly application dependent. In Figure~\ref{fig:cache}, we investigate the effects of caching by evaluating the two possible extremes: the datapoints labeled ``warm'' correspond to a tight loop in which large portions of the index structure and underlying data can be cached between lookups. The datapoints labeled ``cold'' correspond to the same workload, but with additionally fully flushing the cache after each lookup. The gain from a warm cache for all five index structures ranges from 2x to 2.5x. With small index sizes ($<$ 1MB), the cold-cache variant of several learned index structures outperform the warm-cache BTree. With larger (and arguably more realistic) index structure sizes, obviously whether or not the cache is warm or cold is more important than the choice of index structure. Regardless of if the cache is warm or cold, we found that learned approaches exhibited dominant performance / size tradeoffs.


\subsubsection{Memory fences}
\label{sec:fence}

\begin{figure*}
  \includegraphics[width=\textwidth]{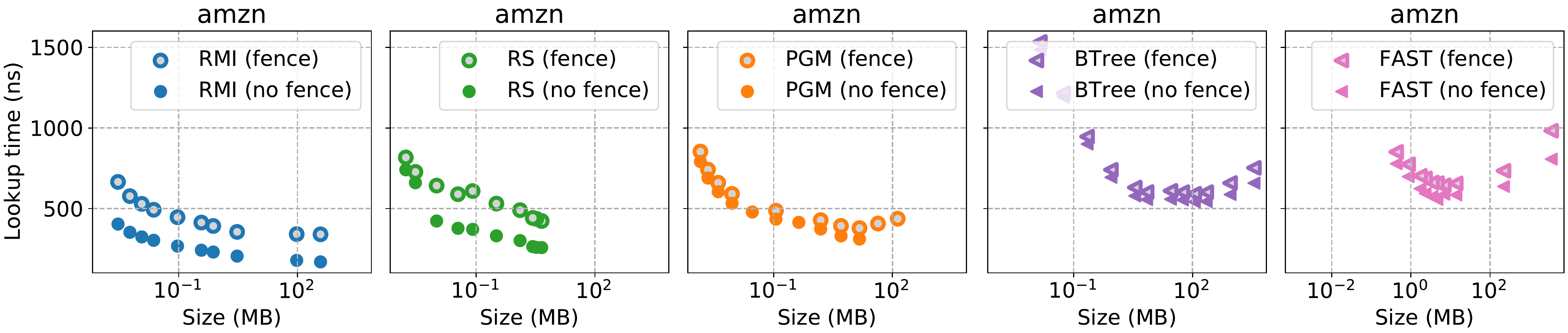}
  \caption{Performance of various index structures with and without a memory fence. Without the fence, the CPU may reorder instructions and overlap computation between lookups. With the fence, each lookup must be completed before the next lookup begins. Extended plots with all techniques and datasets are available here: \url{https://rm.cab/lis4}}
  \label{fig:fence}
\end{figure*}

Modern CPUs and compilers may reorder instructions to overlap computation and memory access or otherwise improve pipelining. For example, consider a simple program that loads $x$, does a computation $f(x)$, loads $y$, and then does a computation $g(y)$. Assuming the load of $y$ does not depend on $x$, a load of $y$ may be reordered to occur before the computation of $f(x)$, so that the latency from loading $y$ can be hidden within the computation of $f(x)$. 
When considering index structures, lookups placed in a tight loop may cause the CPU or compiler to overlap the final computation of one query with the initial memory read of the next query. In some applications, this may be realistic and desirable -- in other applications, expensive computations between index lookups may prevent such overlapping.
Thus, some indexes may disproportionately benefit from this reordering.

To test the impact of reordering on lookup time, we inserted a memory fence 
instruction into our experimental loop. This prevents the CPU or compiler from reordering operations across the fence. Figure~\ref{fig:fence} shows that RMI and RS -- two of the most competitive index structures -- have the largest drop in performance when a memory fence is introduced (approximately a 50\% slowdown). The BTree, FAST and PGM are almost entirely unaffected. While the inclusion of a memory fence harms the performance of RMI and RS, learned structures still provide a better performance / size tradeoff for the \books dataset (results for other datasets are similar, but omitted due to space constraints).

The impact of a memory fence was highly correlated with the number of instructions used by an index structure (Figure~\ref{fig:metrics}): indexes using fewer instructions, like RMI and RS, were impacted to a greater extent than structures using more instructions, like BTrees. Since reordering optimizations often examine only a small window of instructions (i.e., ``peephole optimizations''~\cite{peephole}), reordering optimizations may be more effective when instruction counts are lower. This may explain why RMI and RS are impacted more by a memory fence.

We recommend that future researchers test their index structures with memory fences to determine how much benefit their structure gets from reordering. Getting a lot of benefit from reordering is not necessarily a bad thing; plenty of applications require performing index lookups in a tight loop, with only minimal computation being performed on each result. Ideally, researchers should evaluate their index structures within a specific application, although this is much more difficult.


\subsection{Multithreading}
\label{sec:multithreading}
\begin{figure*}
\centering
\begin{subfigure}{0.32\textwidth}
  \centering
  \includegraphics[width=\textwidth]{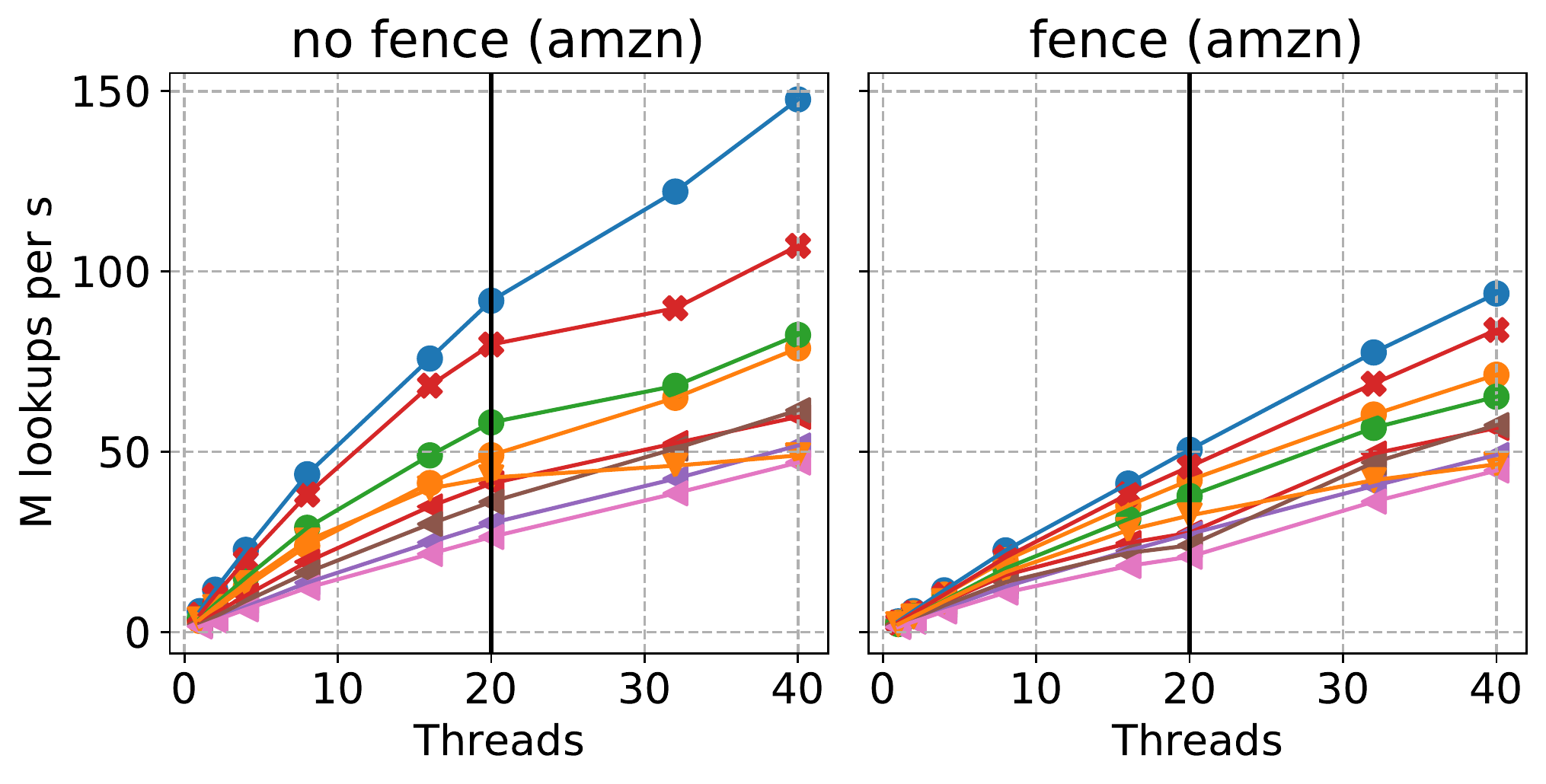}
  \caption{Multithreaded throughput for the \books dataset, models have a fixed size of 50MB. No memory fence (left) and with memory fence (right).}
  \label{fig:threads}
\end{subfigure}\hspace{0.019\textwidth}%
\begin{subfigure}{0.32\textwidth}
  \centering
  \includegraphics[width=\linewidth]{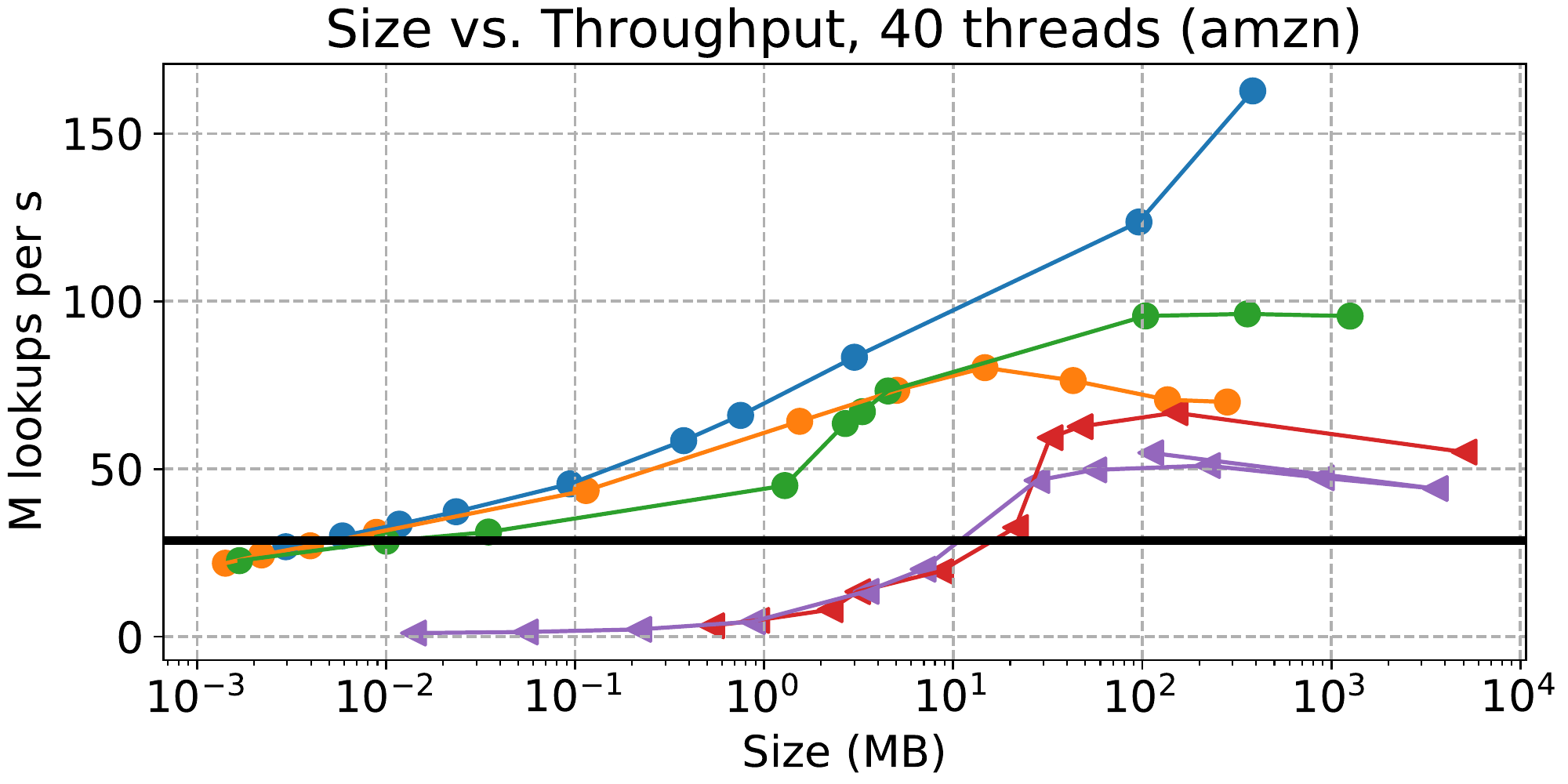}
  \caption{Model size vs. 40-thread throughput for the \books dataset. An extended plot with all index techniques is available here: \url{https://rm.cab/lis6}}
  \label{fig:threads40}
\end{subfigure}\hspace{0.019\textwidth}%
\begin{subfigure}{0.32\textwidth}
  \centering
  \includegraphics[width=\textwidth]{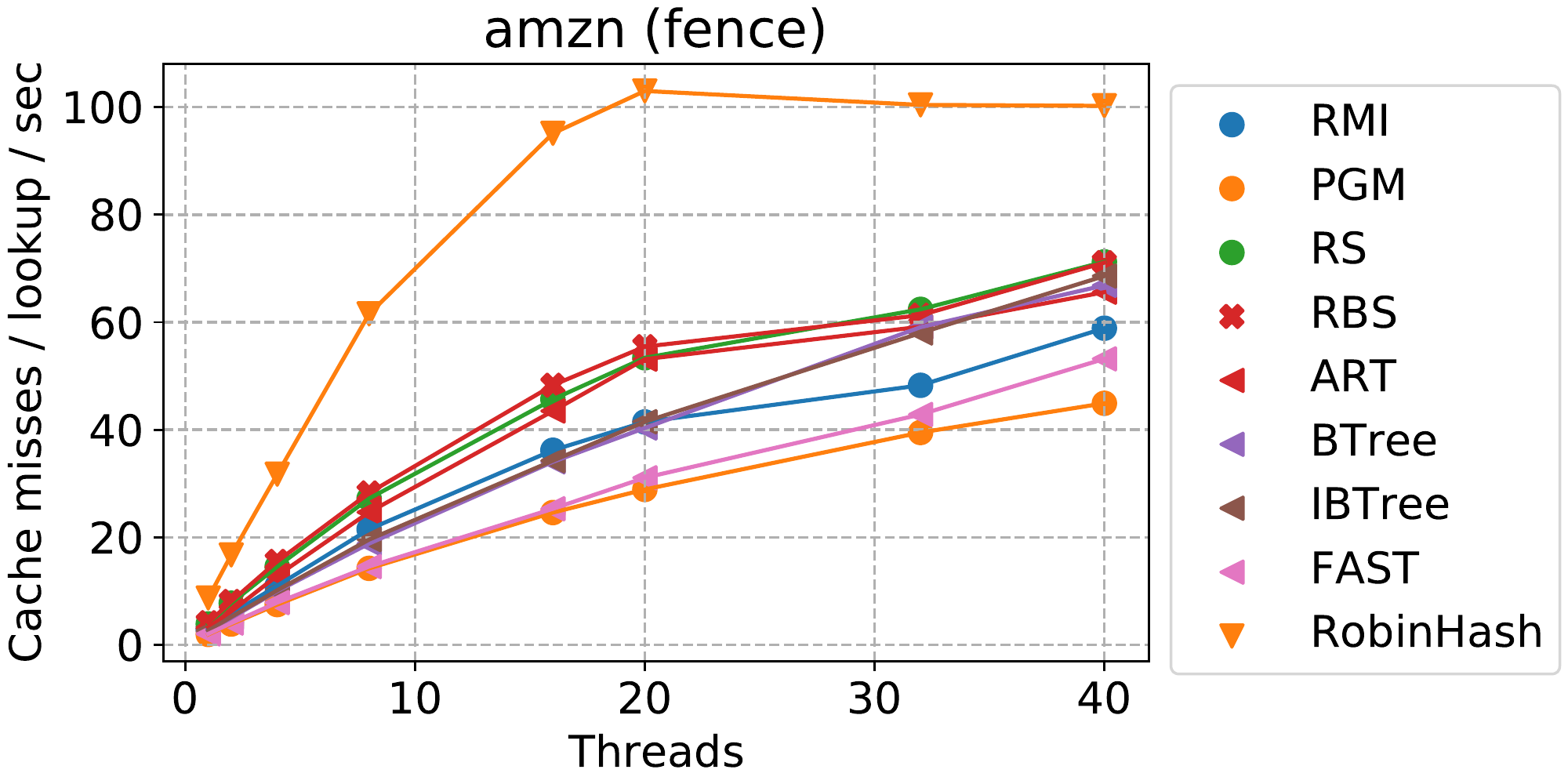}
  \caption{Cache misses per lookup per second for various data structures. More cache misses per second indicates that speedup from multithreading may be negatively impacted.}
  \label{fig:cmps}
\end{subfigure}
\caption{Multithreading results}
\end{figure*}

Here, we evaluate how various index structures scale when queried by concurrent threads. Our test CPU had 20 physical cores, capable of executing 40 simultaneous threads with hyperthreading. Since multithreading strictly increases latency, here we evaluate throughput (lookups per second). 

\sparagraph{Varying thread count} We first vary the number of threads, fixing the model size at 50MB 
except for RobinHash, which is still the full size. The results are plotted in Figure~\ref{fig:threads}, with and without a memory fence. Overall, all three learned index variants scale with an increasing number of threads, although only the RMI achieved higher throughput than the RBS lookup table in this experiment.

RobinHash, the technique with the lowest latency with a single thread, does not achieve the highest throughput in a concurrent environment.\footnote{\small The SIMD Cuckoo implementation~\cite{url-simd-cuckoo} only supports 32-bit keys, and was not included in this experiment.} 
Even the RBS lookup table achieves higher throughput than RobinHash, regardless of whether or not a memory fence was used.
We do not consider hash tables optimized for concurrent environments~\cite{hashing_eval}; here we only demonstrate that an off-the-shelf hash table with a load factor optimized for single-threaded lookups does not scale seamlessly.


To help explain why certain indexes scaled better than others, we measured the number of cache misses incurred \emph{per second} by each structure, plotted in Figure~\ref{fig:cmps}. If a index structure incurs more cache misses per second, then the benefits of multithreading will be diminished, since threads will be latency bound waiting for access to RAM. 
Indeed, RobinHash incurs a much larger number of cache misses per second than any other technique. The larger size of the hash table may contribute to this, as fewer cache lines may be shared in between lookups compared with a smaller index.


PGM and FAST have the fewest cache misses per second at 40 threads, suggesting that PGM and FAST may benefit the most from multithreading. To investigate this, we tabulated the \emph{relative} speedup factor of each technique. Due to space constraints, the plot is available online: \url{https://rm.cab/lis8}. FAST has the highest relative speedup, achieving 32x throughput with 40 threads. In addition to having few cache misses per second, FAST also takes advantage of streaming AVX-512 instructions, which allows for effective overlap of computation with memory reads. 
PGM, despite having the least cache misses per second, achieved only a 27x speedup at 40 threads. On the other hand, RobinHash had by far the most cache misses per second and the lowest relative speedup at 40 threads (20x). Thus, cache misses per second correlate with, but do not always determine, the speedup factor of an index structure. 


\sparagraph{Varying index size} Next, we fix the number of threads at 40, and vary the size of the index. Results are plotted in Figure~\ref{fig:threads40}. One might expect smaller structures to have better throughput because of caching effects; we did not find this to be the case. In general, larger indexes had higher throughput than smaller ones. One possible explanation of this behavior is that smaller models, while more likely to remain cached, produce larger search bounds, which cause more cache misses during the last mile search. 

PGM, BTree, RS, and ART indexes suffered decreased throughput at large model sizes. This suggests that the cache misses incurred from the larger model sizes are not enough to make up for the refinement in the search bound. The RMI did not suffer such a regression, possibly because each RMI inference requires at most two cache misses (one for each model level), whereas for other indexes the number of cache misses per inference could be higher. 

\subsection{Build times}
\label{sec:build}
\begin{figure}
  \includegraphics[width=\linewidth]{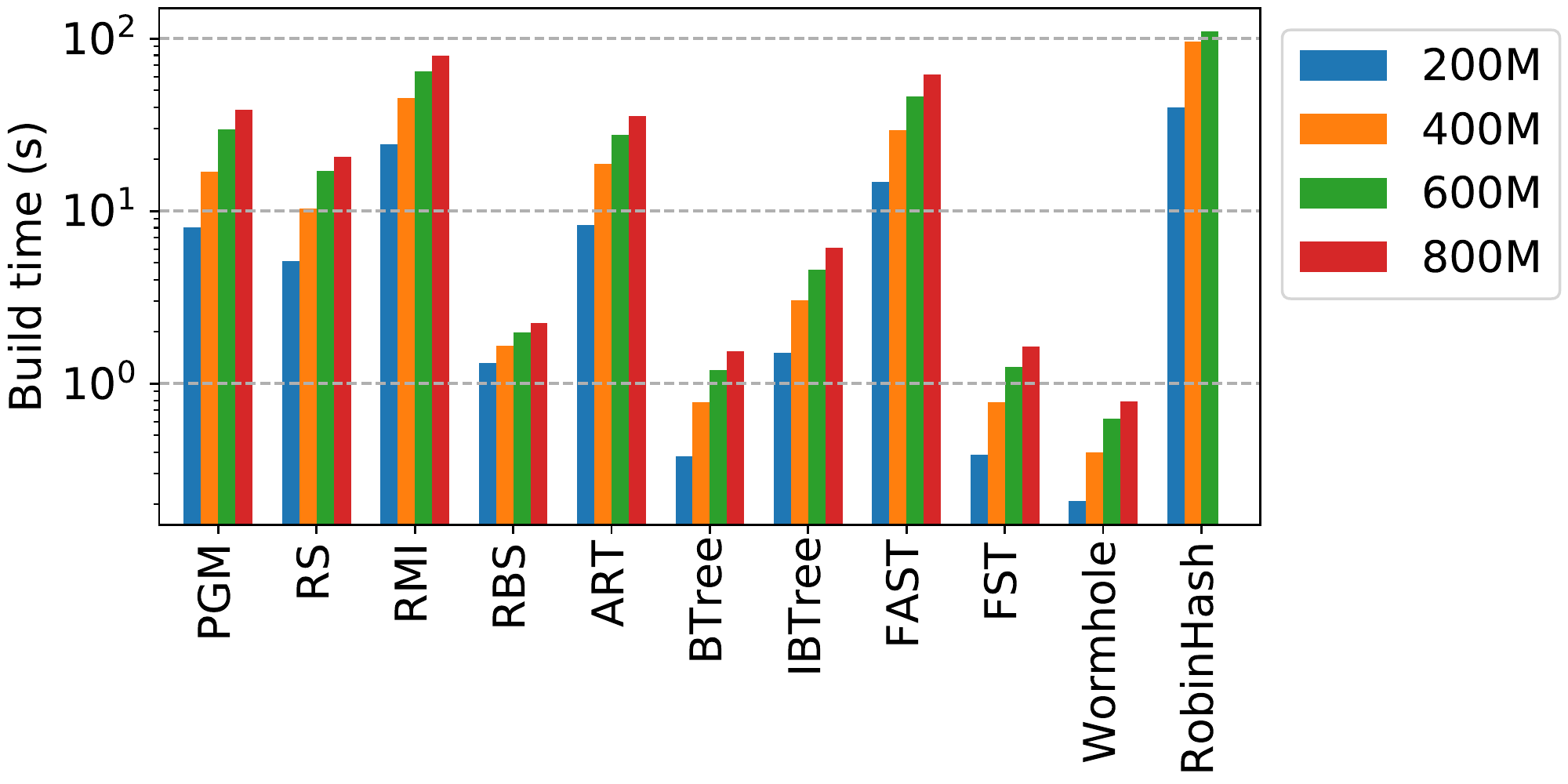}
  \caption{Build times for the fastest (in terms of query time) variant of each index type for the \books dataset at four different data sizes. Note the log scale.}
  \label{fig:build_time_fastest}
\end{figure}

Figure~\ref{fig:build_time_fastest} shows the single-threaded build time required for the fastest (in terms of lookup time) variants of each index structure on \books at different dataset sizes. We do not include the time required to tune each structure (automatically via CDFShop~\cite{cdfshop} for RMIs, manually for other structures). We note that automatically tuning an RMI may take several minutes.
Unsurprisingly, BTrees, FST, and Wormhole provide the fastest build times, as these structures were designed to support fast updates.\footnote{\small In particular, Wormhole and PGM can handle parallel inserts and builds respectively, which we do not evaluate here.} Of the non-learned indexes, FAST and RobinHash have the longest build times. Maximizing the performance of Robinhood hashing requires using a high load factor (to keep the structure compact), which induces a high number of swaps. We note that many variants of Robinhood hashing support parallel operations, and thus lower build times.

For the largest dataset, the build times for the fastest variants of RMI, PGM, and RS were 80 seconds, 38 seconds, and 20 seconds respectively.
Of the learned index structures, RS consistently provides the fastest build times regardless of dataset size. This is explained by the fact that an RS index can be built in a single pass over the data with constant time per element~\cite{radix-spline}. In contrast, while a PGM index could theoretically be built in a single pass, the tested implementation of the PGM index builds the initial layer of the index in a single pass, and builds subsequent layers in a single pass over the previous layer (each logarithmically smaller). RMIs require one full pass over the underlying data per layer. In our experiments, no learned index takes advantage of parallelism during construction, which could provide a speedup.

\section{Conclusion and future work}
\label{sec:conclusion}

In this work, we present an open source benchmark that includes several state-of-the-art tuned implementations of learned and traditional index structures, as well as several real-world datasets. Our experiments on read-only in-memory workloads searching over dense arrays showed that learned structures provided Pareto dominant performance / size behavior. This dominance, while sometimes diminished, persists even when varying dataset sizes, key sizes, memory fences, cold caches, and multi-threading. We demonstrate that the performance of learned index structures is not attributable to any specific metric, although cache misses played the largest explanatory role. In our experiments, learned structures generally had higher build times than insert-optimized traditional structures like BTrees. Amongst learned structures, we found that RMIs provided the strongest performance / size but the longest build times, whereas both RS and PGM indexes could be constructed faster but had slightly slower lookup times.

In the future, we plan to examine the end-to-end impact of learned index structures on real applications. Opportunities to combine a simple radix table with an RMI structure (or vice versa) are also worth investigating. As more learned index structures begin to support updates~\cite{fiting_tree, pgm-index, alex}, a benchmark against traditional indexes (which are often optimized for updates) could be fruitful.

\section*{Acknowledgments}
This research is supported by Google, Intel, and Microsoft as part of the MIT Data Systems and AI Lab (DSAIL) at MIT, NSF IIS 1900933, DARPA Award 16-43-D3M-FP040, and the MIT Air Force Artificial Intelligence Innovation Accelerator (AIIA). 





\end{sloppypar}

\bibliographystyle{abbrv}
\bibliography{main}

\end{document}